\documentclass[11pt]{article}

\usepackage{graphics,graphicx,color,float, hyperref}
\usepackage{epsfig,bbm}
\usepackage{bm}
\usepackage{mathtools}
\usepackage{amsmath}
\usepackage{amssymb}
\usepackage{amsfonts}
\usepackage{amsthm}
\usepackage {times}
\usepackage{layout}
\usepackage{setspace}
\usepackage{geometry}
\usepackage{tikz-cd}

\newtheorem{fact}{Fact}

\newcommand{\beq}{\begin{equation}}
\newcommand{\enq}{\end{equation}}
\newcommand{\bel}{\begin{lemma}}
\newcommand{\enl}{\end{lemma}}
\newcommand{\bet}{\begin{theorem}}
\newcommand{\ent}{\end{theorem}}

\newcommand{\Tr}{\mathrm{Tr}}
\newcommand{\ketbra}[1]{|#1\rangle\langle#1|}

\newcommand{\eps}{\varepsilon}

\newcommand {\varrelent} [2] {\ensuremath{\mathrm{V}{(#1 \| #2)}}}
\newcommand {\hmin} [2] {\fn{\mathrm{H}_{\min}}{#1 \middle | #2}}
\newcommand {\hmineps} [3] {\fn{\mathrm{H}^{#3}_{\min}}{#1 \middle | #2}}
\newcommand {\hmaxeps} [3] {\fn{\mathrm{H}^{#3}_{\max}}{#1 \middle | #2}}
 
\newcommand*{\cL}{\mathcal{L}}
\newcommand*{\cC}{\mathcal{C}}
\newcommand*{\cA}{\mathcal{A}}
\newcommand*{\cR}{\mathcal{R}}
\newcommand*{\cH}{\mathcal{H}}
\newcommand*{\cM}{\mathcal{M}}

\newcommand*{\cB}{\mathcal{B}}
\newcommand*{\cD}{\mathcal{D}}

\newcommand*{\cN}{\mathcal{N}}

\newcommand*{\cE}{\mathcal{E}}

\newcommand{\cP}{\mathcal{P}}

\newcommand{\supp}{\mathrm{supp}}
\newcommand{\suppress}[1]{}

\newcommand{\defeq}{\ensuremath{ \stackrel{\mathrm{def}}{=} }}
\newcommand{\F}{\mathrm{F}}
\newcommand{\Pur}{\mathrm{P}}
\newcommand {\br} [1] {\ensuremath{ \left( #1 \right) }}

\newcommand {\minusspace} {\: \! \!}
\newcommand {\smallspace} {\: \!}
\newcommand {\fn} [2] {\ensuremath{ #1 \minusspace \br{ #2 } }}
\newcommand {\ball} [2] {\fn{\mathcal{B}^{#1}}{#2}}
\newcommand {\relent} [2] {\fn{\mathrm{D}}{#1 \middle\| #2}}
\newcommand {\dmax} [2] {\fn{\mathrm{D}_{\max}}{#1 \middle\| #2}}
\newcommand {\dmaxeps} [3] {\fn{\mathrm{D}^{#3}_{\max}}{#1 \middle\| #2}}
\newcommand {\dheps} [3] {\fn{\mathrm{D}^{#3}_{\mathrm{H}}}{#1 \middle\| #2}}
\newcommand {\mutinf} [2] {\fn{\mathrm{I}}{#1 \smallspace : \smallspace #2}}
\newcommand {\imax}{\ensuremath{\mathrm{I}_{\max}}}
\newcommand {\imaxeps}{\ensuremath{\mathrm{I}^{\varepsilon}_{\max}}}
\newcommand {\condmutinf} [3] {\mutinf{#1}{#2 \smallspace \middle\vert \smallspace #3}}
\newcommand {\id}{\mathrm{I}}
\newcommand {\bsigma}{\bar{\sigma}}
\newcommand {\barq}{\bar{q}}

\newcommand{\bra}[1]{\langle #1|}
\newcommand{\ket}[1]{|#1 \rangle}

\mathchardef\mhyphen="2D

\makeatletter
\newcommand*{\rom}[1]{\expandafter\@slowromancap\romannumeral #1@}
\makeatother

\mathchardef\mhyphen="2D

\newtheorem{remark}{Remark}
\newtheorem{definition}{Definition}
\newtheorem{claim}{Claim}

\newtheorem{theorem}{Theorem}
\newtheorem{lemma}{Lemma}
\newtheorem{corollary}{Corollary}

\begin {document}
\begin{titlepage}
\title{Convex-split and hypothesis testing approach to one-shot quantum measurement compression and randomness extraction}
\author{
Anurag Anshu\footnote{Centre for Quantum Technologies, National University of Singapore, Singapore. \texttt{a0109169@u.nus.edu}} \qquad
Rahul Jain\footnote{Centre for Quantum Technologies, National University of Singapore; MajuLab, UMI 3654, 
Singapore and VAJRA Adjunct Faculty, TIFR, Mumbai, India. \texttt{rahul@comp.nus.edu.sg}} \qquad 
Naqueeb Ahmad Warsi\footnote{Centre for Quantum Technologies, National University of Singapore and School of Physical and Mathematical Sciences, Nanyang Technological University, Singapore and IIITD, Delhi. \texttt{warsi.naqueeb@gmail.com}} 
}\date{}

\end{titlepage}
\maketitle

\abstract{We consider the problem of {\em quantum measurement compression} with side information in the one-shot setting with shared randomness. In this problem, Alice shares a pure state with Reference and Bob and she performs a measurement on her registers. She wishes to communicate the outcome of this measurement to Bob using shared randomness and classical communication, in such a way that the outcome that Bob receives is correctly correlated with Reference and Bob's own registers. Our goal is to simultaneously minimize the classical communication and randomness cost. We provide a protocol based on {\em convex-split} and {\em position based decoding} with its communication upper bounded in terms of smooth max and hypothesis testing relative entropies. 

We also study the randomness cost of our protocol in both one-shot and asymptotic and i.i.d. setting. By generalizing the convex-split technique to incorporate pair-wise independent random variables, we show that our one shot protocol requires small number of bits of shared randomness. This allows us to construct a new protocol in the asymptotic and i.i.d. setting, which is optimal in both the number of bits of communication and the number of bits of shared randomness required.

We construct a new protocol for the task of strong randomness extraction in the presence of quantum side information. Our protocol achieves error guarantee in terms of relative entropy (as opposed to trace distance) and extracts close to optimal number of uniform bits. As an application, we provide new achievability result for the task of quantum measurement compression without feedback, in which Alice does not need to know the outcome of the measurement. This leads to the optimal number of bits communicated and number of bits of shared randomness required, for this task in the asymptotic and i.i.d. setting.
}

\section{Introduction}
The formalism of quantum mechanics is well known to be statistical in nature, which limits an experimenter's knowledge about a given quantum system. Quantum measurement serves as the tool for obtaining this statistical information, which can be used for further physical or information theoretic operations on the system. In fact, a large part of quantum information theory is about finding most suitable quantum measurements in a given scenario, such as for distinguishing quantum states or designing quantum algorithms.  In this backdrop, an elementary but fundamentally important problem is to understand how much information does a measurement statistic reveal about a quantum system. 

This problem was given a firm information theoretic treatment in the seminal work by Winter~\cite{Winter04}, building upon the ideas developed in \cite{MassarP00} and the follow-up work~\cite{WinterM01}. Consider the setting where Alice and Reference share $n$ copies of a joint pure state $\ket{\Psi}_{RA}$ and Alice wishes to communicate to Bob the outcome of a quantum measurement or POVM $\Lambda$ (which is a collection $\{\Lambda_c\}$ of positive operators such that $\sum_c \Lambda_c = \id$) performed on her registers $A^n$. It was shown in \cite{Winter04} that with the aid of shared randomness, the amount of classical communication required by Alice is the mutual information between Reference and measurement outcomes. This was achieved by showing that instead of performing the measurement $\Lambda$ itself, Alice could consider a decomposition of $\Lambda$ in terms of a convex combination of POVMs $\{\Lambda^j\}$ and send the outcome of the measurement $\Lambda^j$ on her registers conditioned on sampling $j$ from shared randomness. 

The work has found important applications in several information theoretic tasks (such as in \cite{HsiehW15}) and for distilling pure states from bi-partite mixed states \cite{Devetak05,HorodeckiOSS03,HorodeckiOSSS05,DevetakKrovi07, DevetakHW08}. Subsequently its extension with quantum side information was considered by Wilde, Hayden, Buscemi and Hsieh ~\cite{WildeHBH12} in the asymptotic setting. Here, Alice, Bob and Reference share a joint pure state and Alice wishes to transmit the measurement results to Bob. One can expect further compression in the communication due to the side information with Bob, which was shown to hold in ~\cite{WildeHBH12}. This work also provides a detailed overview of the result in \cite{Winter04} and discusses several related scenarios. 

\begin{figure}[ht]
\centering
\begin{tikzpicture}[xscale=0.9,yscale=1.1]

\draw[ultra thick, fill=blue!10!white] (-6,6) rectangle (12,1);

\draw[thick, fill=blue!40!white] (-2.8,5) -- (-4.3,3) -- (-1.3,2.5) -- (-2.8,5);
\node at (-2.5,5) {$R$};
\node at (-4.5,3) {$A$};
\node at (-1.1,2.5) {$B$};
\node at (-2.8,3.5) {$\ket{\Psi^0}_{RAB}$};

\node at (-3.0,5.6) {Reference};
\node at (-4.5,4.6) {Alice};
\node at (-0.8,4.6) {Bob};

\draw[->, thick] (-0.3,3.5) -- (1.2,3.5);
\node at (0.4, 3.8) {Alice};
\node at (0.4, 3.2) {measures};


\draw[thick, fill=blue!40!white] (3.2,5) -- (1.7,3) --  (4.7,2.5) -- (3.2,5);
\draw[thick, fill=blue!40!white] (1.7,2.2) rectangle (2.5, 2.5);
\node at (3.5,5) {$R$};
\node at (1.5,3) {$A$};
\node at (1.5,2) {$\bar{C}$};
\node at (-2.7,2) {$C$};
\node at (4.9,2.5) {$B$};
\node at (3.2,3.5) {$\ket{\Psi}^c_{RAB}$};
\node at (2.1,2.33) {$c$};

\node at (3,5.6) {Reference};
\node at (1.5,4.6) {Alice};
\node at (4.5,4.6) {Bob};

\draw[->, thick] (5.7,3.5) -- (7.2,3.5);

\draw[thick, fill=blue!40!white] (9.2,5) -- (7.7,2.5) -- (10.7,3) -- (9.2,5);
\draw[thick, fill=blue!40!white] (7.7,2.2) rectangle (10.7,1.8);
\node at (9.5,5) {$R$};
\node at (7.5,2.5) {$A$};
\node at (10.9,3) {$B$};
\node at (10.7,1.6) {$C$};
\node at (9.5,3.5) {$\ket{\Psi}^c_{RAB}$};
\node at (7.9,2) {$c$};
\node at (10.5,2) {$c$};

\node at (9,5.6) {Reference};
\node at (7.5,4.6) {Alice};
\node at (10.5,4.6) {Bob};

\end{tikzpicture}
\caption{\small Quantum measurement compression task. Alice applies a measurement on her register to obtain measurement outcome in register $C$. Her task is to communicate the measurement outcome to Bob with the aid of only shared randomness.}
 \label{fig:measurementcompress}
\end{figure}
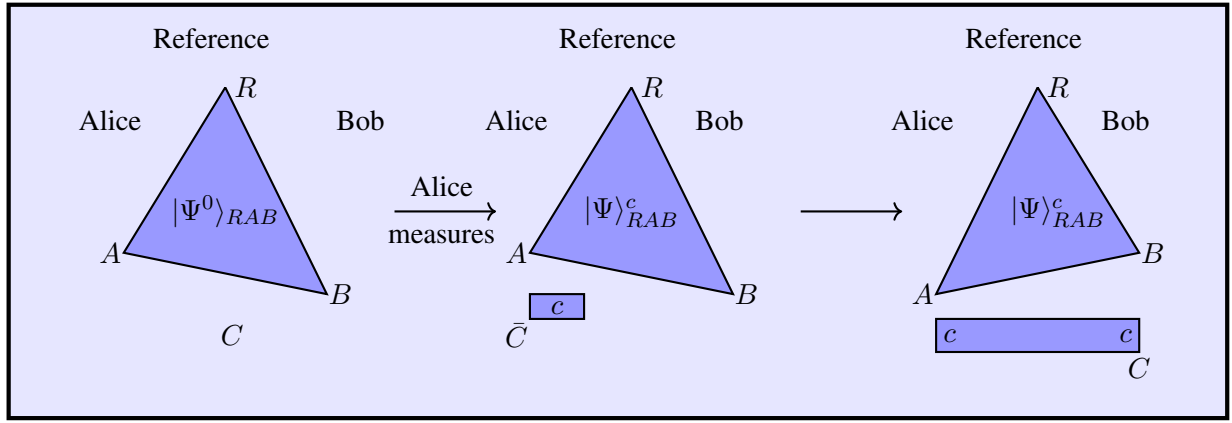

We consider the same problem in the one shot setting. One-shot information theory provides a framework for information processing in the scenarios which go beyond asymptotic and i.i.d. Apart from being relevant for practical scenarios, this framework also provides insights into the inner workings of information protocols, as the complications (and conveniences) arising due to many copies of the state are no longer present. Many quantum tasks have been formulated in their one-shot setting, such as quantum state merging (\cite{Berta09, Renner11}, originally introduced in \cite{horodecki07}) and quantum state redistribution (\cite{Berta14,AnshuDJ14,AnshuJW17-1}, originally introduced in \cite{Devatakyard, YardD09}). In this setting, the task of quantum measurement compression is as follows (Figure \ref{fig:measurementcompress})

\vspace{0.1in}

\noindent {\bf Quantum measurement compression task:} Alice ($A$), Bob ($B$) and Reference ($R$) share a joint pure state $\ket{\Psi^0}_{RAB}$. Alice performs a measurement on her register $A$, described by the POVM $\Lambda$ with POVM elements $\{\Lambda_c\}_c$. Since Alice also generates the record of the measurement in a register $C$, the overall transformation on the shared state can be viewed as 
$$\ketbra{\Psi^0}_{RAB} \rightarrow \Psi_{RAB}:=\sum_c p(c) \ketbra{\psi^c}_{RAB}\otimes \ketbra{c}_C,$$ where $\ketbra{\psi^c}$ is the \textit{post-selected state} on the measurement outcome $c$ and $p(c)$ is the probability of this outcome. An equivalent way of phrasing this is as follows, which shall be crucial in our analysis. Alice attaches ancilla registers $C\bar{C}$ in a standard state and performs a unitary on her side to produce the following state:
$$ \ket{\Psi}_{RAC\bar{C}B} = \sum_c \sqrt{p(c)} \ket{c}_C \ket{c}_{\bar{C}} \ket{\psi^c}_{RAB} .$$
Upon tracing out the register $\bar{C}$, Alice recovers the desired post-measurement state. The objective is that using \textit{shared randomness} and \textit{classical communication}, Alice should communicate register $C$ to Bob. In other words, Bob should produce a register $C'$ such that the state in registers $RABCC'$ after the protocol is $\Phi_{RABCC'}$ satisfying 
$$\Pur(\Phi_{RABCC'}, \sum_c p(c) \ketbra{c}_C\otimes\ketbra{c}_{C'} \otimes \ketbra{\psi^c}_{RAB})\leq \eps , $$
where $\eps > 0$ is error parameter and $\Pur(\cdot, \cdot)$ represents purified distance. We note that the register $\bar{C}$ is not taken into account in the final state. This makes the present task different from the task of quantum state redistribution. In some sense, present task is a hybrid between classical source coding and quantum state redistribution.

The work~\cite{WildeHBH12} gave the optimal communication rate required to achieve this task in the asymptotic and i.i.d. setting, showing that the communication rate is equal to $\condmutinf{R}{C}{B}_{\Psi}$ and showed that the number of bits of shared randomness required in the protocol is $\mathrm{H}(C|RB)_{\Psi}$. A related result that involved sending classical message in presence of quantum side information in the one-shot setting appeared in the work \cite{RenesR12}.

A natural variant of above task is when the correctness criteria is weakened to 
$$\Pur(\Phi_{RBC'}, \sum_c p(c)\ketbra{c}_{C'} \otimes \ketbra{\psi^c}_{RB})\leq \eps.$$ 
That is, Alice does not need to hold the outcome of the measurement. This task is known as quantum measurement compression without feedback, and was first studied in~\cite{WildeHBH12}. The communication rate and randomness required is now characterized by an auxillary random variable $W$, which can be generated by Alice and using which Bob can generate the actual measurement outcome $C$. Thus, it suffices for Alice to generate $W$ by some quantum measurement and then communicate $C$ to Bob. The work~\cite{WildeHBH12} characterized the communication cost and randomness cost of this task in the asymptotic and i.i.d. setting.

\vspace{0.1in}

\noindent {\bf Randomness extraction:} Randomness extraction is a fundamental task in the cryptographic setting, where one is required to extract uniform bits of randomness from a non-uniform source. If there is a party holding side information about the non-uniform source, then it is desirable to have the uniform extracted bits independent of the party. It is well known that some additional amount of randomness is required in the task, which acts in a catalytic way (is returned after use) for strong randomness extractors. The problem of randomness extraction in the presence of quantum side information has been studied in the many works (see, for example, \cite{Renner05, KonigT08, Tashma09, TomRSS10, Hayashi11, DePVR12, Berta13, BertaFW14}) and is closely related to various other cryptographic primitives such as privacy amplification \cite{BennettBCM95} and information reconciliation \cite{Renner05}. Performance of a randomness extractor is measured in terms of the number of uniform bits that are extracted and the number of bits of additional randomness required.

\vspace{0.1in}

\noindent{\bf Our results:} We consider the task of quantum measurement compression in the one-shot setting (Section \ref{sec:quantumstate}) and present a protocol with communication upper bounded by
$$\min_{\sigma_C}\bigg(\dmaxeps{\Psi_{RBC}}{\Psi_{RB}\otimes \sigma_C}{\eps} - \mathrm{D}_{\mathrm{H}}^{\eps^2}(\Psi_{BC}\|\Psi_{B}\otimes \sigma_C) \bigg)+ \mathcal{O}\left(\log\left(\frac{1}{\varepsilon}\right)\right) ,$$ where $\sigma_C$ is a classical state (that is, it commutes with $\Psi_C$). Above, $\dmaxeps{.}{.}{\eps}$ is the smooth max-relative entropy and $\mathrm{D}_{\mathrm{H}}^{\eps^2}(.\|.)$ is the quantum hypothesis testing relative entropy. We note that this bound converges to $\condmutinf{R}{C}{B}_{\Psi}$ in the asymptotic and i.i.d. setting. A one-shot converse bound for this task appears in~\cite[Lemma 4.1]{LeditzkyWD16}, in terms of the R\'{e}nyi conditional entropies. We provide a converse result in Section \ref{subsec:convbound} in terms of smooth max-relative entropy, closely following the converse results given in \cite{Berta14}.

We also consider the shared randomness cost of the protocol. We show that the number of bits of shared randomness consumed by the protocol is given by $-\dmaxeps{\Psi_{RBC}}{\Psi_{RB}\otimes \id_C}{\eps}$, a one shot analogue of the conditional entropy $\mathrm{H}(C|RB)_{\Psi}$. While our one-shot protocol also requires a small amount of extra randomness to begin with (which is approximately $\log|C|$ bits), this randomness is returned with high fidelity. By reusing it, we find that the rate of shared randomness required in the asymptotic and i.i.d. setting is equal to $\mathrm{H}(C|RB)_{\Psi}$. This, thus recovers the results in \cite{WildeHBH12}, with a conceptually different proof. 

Further, we provide a new protocol for (strong) randomness extraction in presence of quantum side information (Section \ref{sec:randext}). Our high level idea is that randomness extraction and quantum measurement compression must be closely related, as in the former case uniform random bits are gained (that are independent of other quantum systems), whereas in the latter case, the random bits are consumed. Our protocol for randomness extraction follows from the convex-split lemma based on pairwise independence (to be discussed below and obtained in Section \ref{sec:convsplit}). The number of uniform bits extracted and the number of bits of initial randomness required are similar to the `pairwise independent hash function' based randomness extractor discussed in \cite{Renner05}. That is, the number of uniform bits extracted is approximately the \textit{conditional min-entropy} of the source, and the number of bits of initial randomness is twice the number of bits of the source (which is much larger than the best known constructions \cite{DePVR12}). But we highlight that our construction shifts from the standard paradigm of hash function based extractors. 

An application of our randomness extractor is that, when combined with our protocol for quantum measurement compression, we obtain a one-shot protocol for quantum measurement compression without feedback (Section \ref{sec:randext}). The protocol runs the quantum measurement compression for $W$, and then extracts shared randomness from $W$ that is independent of other relevant registers. We show, in the asymptotic and i.i.d. setting, that the communication cost and the number of bits of shared randomness required match with that obtained in \cite{WildeHBH12} (which is optimal, as already shown in \cite{WildeHBH12}). Again, our protocol is conceptually different, and in fact shows that quantum measurement compression without feedback is a `composition' of quantum measurement compression and randomness extraction.  

\vspace{0.1in}

\noindent {\bf Techniques for the achievability result:} We use the two techniques of convex-split (introduced in \cite{AnshuDJ14}) and position based decoding (introduced in \cite{AnshuJW17}) for our achievability result for quantum measurement compression. As mentioned earlier, the task of quantum measurement compression appears to have a close resemblance to the task of quantum state redistribution where the register to be communicated is classical. However an important difference is that for quantum state redistribution, the shared resource allowed between Alice and Bob is quantum entanglement whereas in quantum measurement compression only classical randomness is allowed as a shared resource. This makes this task a hybrid of classical and quantum state redistribution and requires a careful treatment. 

Appealing to this hybrid setting, we use a special hybrid case of Uhlmann's theorem. In the usual setting Uhlmann's theorem is used for bipartite pure quantum states and there is no version of it for for bipartite mixed quantum states. A reason for this is that in bipartite pure states both the systems carry ``full information'' about each other which is not the case with general mixed states. We consider mixed quantum states where mixed states are {\em classical-pure}, that is a classical mixture of pure quantum states and the classical part appears as a copy in both the systems of the bipartite state. Hence the two systems continue to have `full information' about each other. This hybrid Uhlmann's theorem follows naturally from regular Uhlmann's theorem. Equipped with this version of Uhlmann's theorem, we construct the desired protocol as given in Theorem \ref{newcompression}.

\vspace{0.1in}

\noindent \textbf{Optimizing the randomness cost:} While above two techniques give an optimal communication rate in the asymptotic and i.i.d. setting, they do not give the optimal rate of shared randomness required by the protocol. The issue is that the convex-split lemma uses a large amount of additional quantum states in its statement. We remedy this problem by proving a new statement for the convex-split lemma for classical-quantum states (Lemma \ref{convsplitpairwise}), which is one of the main technical contributions of this work. This statement uses pairwise independent random variables and hence leads to substantial reduction in the randomness cost (exponentially small in comparison to the statement given in \cite{AnshuDJ14}). An interesting aspect of using pairwise independent random variables is that position-based decoding can also be performed on it, without any reduction in the efficiency.

Convex-split lemma has recently been applied for classical-quantum states in the setting of one-shot private quantum capacity in \cite{Wilde17a}. Our new statement implies that the codebook required for this protocol requires only pairwise independent random variables. This considerably simplifies the derandomization task, as the support size of pairwise independent random variables is exponentially smaller than independent random variables. Similar arguments apply for the applications of our techniques given in the work \cite{AnshuJW17_cc}, to various network setting in classical information theory. 

\vspace{0.1in}

\noindent {\bf Connection with Winter's approach:} One of the central techniques used in \cite{Winter04} was that of Operator-Chernoff bound (proved in \cite{AhlswedeW02}), to derive the following inequality (below we give a `one-shot' statement, it was originally stated in asymptotic and i.i.d setting). Let $C_1,C_2\ldots C_n$ be independent and identically distributed random variables such that $C_i\sim p$ and let $\rho_{AC} : = \sum_{c}p(c)\ketbra{c}\otimes \rho^c_A$ be a classical quantum state. Then choosing $n$ large enough (as a function of error parameter $\eps$), we have 
\begin{equation}
\label{eq:avgeq}
\sum_{c_1,c_2\ldots c_n} p(c_1)p(c_2)\ldots p(c_n)\|\frac{1}{n}\rho^{c_1}_A+\frac{1}{n}\rho^{c_2}_A+\ldots \frac{1}{n}\rho^{c_n}_A - \rho_A\|_1 \leq \eps.
\end{equation}
This statement was then used in the construction of the desired decomposition of the measurement operator. We note that Operator-Chernoff bound used by Winter is a stronger statement than above, as it says that the probability that $\frac{1}{n}\rho^{c_1}_A+\frac{1}{n}\rho^{c_2}_A+\ldots \frac{1}{n}\rho^{c_n}_A \notin (1\pm \eps)\rho_A$ decays exponentially in $n\eps^2$.

It is possible to see that convex split technique implies Equation \ref{eq:avgeq} (as discussed in Corollary \ref{coveringlemconvsplit}), leading to a connection between both approaches on a broader level. On the other hand, convex split technique is stronger than Equation \ref{eq:avgeq} as it is applicable to coherent setting as well, of which the classical-quantum setting considered above is a special case.

It is also known that Equation \ref{eq:avgeq} is central in the context of private quantum capacity \cite{Devetak05private} (also known as the quantum wiretap channel). Recently, two different works gave one shot bounds for private capacity of a wiretap channel: the work \cite{Wilde17a} used the convex split technique, whereas the work \cite{RadhaSW17} used extensions of Operator-Chernoff bound. Our discussion above suggests interesting connection between both the approaches.

\section{Quantum information theory}
\label{sec:prelims}

For a finite set $\cC$, a probability distribution is a function $p:\cC \rightarrow [0,1]$ satisfying $\sum_{c\in \cC}p(c)=1$. For a finite set $\cC$ and an integer $n$, the probability distribution $p(c_1,c_2,\ldots c_n)$ on $\cC\times \cC \times \ldots \cC$ is \textit{pairwise independent} if $p(c_i,c_j) = p(c_i)p(c_j)$ for all $i,j \in [n]$.

Consider a finite dimensional Hilbert space $\cH$ endowed with an inner product $\langle \cdot, \cdot \rangle$ (in this paper, we only consider finite dimensional Hilbert-spaces). The $\ell_1$ norm of an operator $X$ on $\cH$ is $\| X\|_1:=\Tr\sqrt{X^{\dagger}X}$ and $\ell_2$ norm is $\| X\|_2:=\sqrt{\Tr XX^{\dagger}}$. A quantum state (or a density matrix or a state) is a positive semi-definite matrix on $\cH$ with trace equal to $1$. It is called {\em pure} if and only if its rank is $1$. A sub-normalized state is a positive semi-definite matrix on $\cH$ with trace less than or equal to $1$. Let $\ket{\psi}$ be a unit vector on $\cH$, that is $\langle \psi,\psi \rangle=1$.  With some abuse of notation, we use $\psi$ to represent the state and also the density matrix $\ketbra{\psi}$, associated with $\ket{\psi}$. Given a quantum state $\rho$ on $\cH$, {\em support of $\rho$}, called $\text{supp}(\rho)$ is the subspace of $\cH$ spanned by all eigen-vectors of $\rho$ with non-zero eigenvalues.
 
A {\em quantum register} $A$ is associated with some Hilbert space $\cH_A$. Define $|A| := \dim(\cH_A)$. Let $\mathcal{L}(A)$ represent the set of all linear operators on $\cH_A$  and $\cP(A)$ represent the set of positive semi-definite operators. We denote by $\mathcal{D}(A)$, the set of quantum states on the Hilbert space $\cH_A$. State $\rho$ with subscript $A$ indicates $\rho_A \in \mathcal{D}(A)$. If two registers $A,B$ are associated with the same Hilbert space, we shall represent the relation by $A\equiv B$.  Composition of two registers $A$ and $B$, denoted $AB$, is associated with Hilbert space $\cH_A \otimes \cH_B$.  For two quantum states $\rho\in \mathcal{D}(A)$ and $\sigma\in \mathcal{D}(B)$, $\rho\otimes\sigma \in \mathcal{D}(AB)$ represents the tensor product (Kronecker product) of $\rho$ and $\sigma$. The identity operator on $\cH_A$ (and associated register $A$) is denoted $I_A$. 

Let $\rho_{AB} \in \mathcal{D}(AB)$. We define
\[ \rho_{B} := \Tr_{A}{\rho_{AB}}
:= \sum_i (\bra{i} \otimes I_{B})
\rho_{AB} (\ket{i} \otimes I_{B}) , \]
where $\{\ket{i}\}_i$ is an orthonormal basis for the Hilbert space $\cH_A$.
The state $\rho_B\in \mathcal{D}(B)$ is referred to as the marginal state of $\rho_{AB}$. Unless otherwise stated, a missing register from subscript in a state will represent partial trace over that register. Given a $\rho_A\in\mathcal{D}(A)$, a {\em purification} of $\rho_A$ is a pure state $\rho_{AB}\in \mathcal{D}(AB)$ such that $\Tr_{B}{\rho_{AB}}=\rho_A$. Purification of a quantum state is not unique. A quantum state $\rho_{AB}$ is \textit{classical-quantum} with $A$ being the classical register, if it is of the form $\rho_{AB}= \sum_a p(a)\ketbra{a}\otimes \rho^a_B$, where $\{\ket{a}\}_a$ forms a basis, $\{p(a)\}_a$ is a probability distribution and $\rho^a_B\in \mathcal{D}(B)$. Given such a classical-quantum state $\rho_{AB}$ with $A$ being the classical register, we shall denote the state on register $B$ conditioned on the value $a$ in register $A$ by $\rho^a_B$. 

A quantum {map} $\cE: \mathcal{L}(A)\rightarrow \mathcal{L}(B)$ is a completely positive and trace preserving (CPTP) linear map (mapping states in $\mathcal{D}(A)$ to states in $\mathcal{D}(B)$). A quantum measurement $\cN: \cL(A) \rightarrow \cL(A'C)$ is characterized by a collection of operators $\{N_c : \cH_A \rightarrow \cH_{A'}\}$ that satisfy $\sum_c N_c^{\dagger}N_c = \id_A$ and is given by
$$\cN(\rho_A) = \sum_c \ketbra{c}_C\otimes N_c\rho_AN^{\dagger}_c.$$
 A {\em unitary} operator $U_A:\cH_A \rightarrow \cH_A$ is such that $U_A^{\dagger}U_A = U_A U_A^{\dagger} = I_A$. An {\em isometry}  $V:\cH_A \rightarrow \cH_B$ is such that $V^{\dagger}V = I_A$ and $VV^{\dagger} = I_B$. The set of all unitary operations on register $A$ is  denoted by $\mathcal{U}(A)$.

We shall consider the following information theoretic quantities. All the logarithms is in base $2$. We consider only normalized states in the definitions below. Let $\varepsilon \in (0,1)$. 
\begin{enumerate}
\item {\bf Fidelity} (\cite{Josza94}, see also \cite{uhlmann76}). For $\rho_A,\sigma_A \in \mathcal{D}(A)$, $$\F(\rho_A,\sigma_A)\defeq\|\sqrt{\rho_A}\sqrt{\sigma_A}\|_1.$$ For classical probability distributions $P = \{p_i\}, Q =\{q_i\}$, $$\F(P,Q)\defeq \sum_i \sqrt{p_i \cdot q_i}.$$  
\item {\bf Trace distance}. For $\rho_A,\sigma_A \in \mathcal{D}(A)$, $$\Delta(\rho_A,\sigma_A)\defeq\frac{1}{2}\|\rho_A - \sigma_A\|_1.$$
\item {\bf Purified distance} (\cite{GilchristLN05}) For $\rho_A,\sigma_A \in \mathcal{D}(A)$, $$\Pur(\rho_A,\sigma_A) \defeq \sqrt{1-\F^2(\rho_A,\sigma_A)}.$$
\item {\bf $\varepsilon$-ball} For $\rho_A\in \mathcal{D}(A)$, $$\ball{\eps}{\rho_A} \defeq \{\rho'_A\in \mathcal{D}(A)|~\Pur(\rho_A,\rho'_A) \leq \varepsilon\}. $$ 

\item {\bf Von Neumann entropy} (\cite{Neumann32}) For $\rho_A\in\mathcal{D}(A)$, $$S(\rho_A) \defeq - \Tr(\rho_A\log\rho_A) .$$ 
\item {\bf Relative entropy} (\cite{umegaki1954}) For $\rho_A,\sigma_A\in \mathcal{D}(A)$ such that $\text{supp}(\rho_A) \subset \text{supp}(\sigma_A)$, $$\relent{\rho_A}{\sigma_A} \defeq \Tr(\rho_A\log\rho_A) - \Tr(\rho_A\log\sigma_A) .$$ 
\item {\bf Relative entropy variance.} For $\rho_A,\sigma_A\in \mathcal{D}(A)$ such that $\text{supp}(\rho_A) \subset \text{supp}(\sigma_A)$,
$$\varrelent{\rho}{\sigma} := \Tr(\rho(\log\rho - \log\sigma)^2) - (\relent{\rho}{\sigma})^2.$$
\item {\bf Mutual information} For $\rho_{AB}\in \mathcal{D}(AB)$, $$\mutinf{A}{B}_{\rho}\defeq S(\rho_A) + S(\rho_B)-S(\rho_{AB}) = \relent{\rho_{AB}}{\rho_A\otimes\rho_B}.$$
\item {\bf Conditional mutual information} For $\rho_{ABC}\in\mathcal{D}(ABC)$, $$\condmutinf{A}{B}{C}_{\rho}\defeq \mutinf{A}{BC}_{\rho}-\mutinf{A}{C}_{\rho}.$$
\item {\bf Conditional entropy} For $\rho_{AB}\in \mathcal{D}(AB)$, $$\mathrm{H}(A|B) = S(\rho_{AB}) - S(\rho_B).$$
\item {\bf Max-relative entropy} (\cite{Datta09}) For $\rho_A\in \mathcal{D}(A)$ and $\sigma_A\in \cP(A)$ such that $\text{supp}(\rho_A) \subset \text{supp}(\sigma_A)$, $$ \dmax{\rho_A}{\sigma_A}  \defeq  \inf \{ \lambda \in \mathbb{R} : 2^{\lambda} \sigma_A \geq \rho_A \}  .$$  
\item {\bf Smooth max-relative entropy} (\cite{Datta09}, see also \cite{Jain:2009}) For $\rho_A\in \mathcal{D}(A)$ and $\sigma_A\in \cP(A)$ such that $\text{supp}(\rho_A) \subset \text{supp}(\sigma_A)$, $$ \dmaxeps{\rho_A}{\sigma_A}{\eps}  \defeq  \inf_{\rho'_A \in \ball{\eps}{\rho_A}} \dmax{\rho'_A}{\sigma_A} .$$  
\item {\bf Hypothesis testing relative entropy} (\cite{BuscemiD10}, see also \cite{hayashinagaoka})  For $\rho_A,\sigma_A\in \mathcal{D}(A)$, $$ \dheps{\rho_A}{\sigma_A}{\eps}  \defeq  \sup_{0<\Pi<I, \Tr(\Pi\rho_A)\geq 1-\eps}\log\left(\frac{1}{\Tr(\Pi\sigma_A)}\right).$$  
\item {\bf Conditional min-entropy} (\cite{Renner05}) For $\rho_{AB}\in \mathcal{D}(AB)$, $$ \hmin{A}{B}_{\rho} \defeq  - \inf_{\sigma_B\in \mathcal{D}(B)}\dmax{\rho_{AB}}{\id_{A}\otimes\sigma_{B}} .$$  	
\item {\bf Smooth conditional min-entropy} (\cite{Renner05}) For $\rho_{AB}\in \mathcal{D}(AB)$, $$\hmineps{A}{B}{\eps}_{\rho} \defeq   \sup_{\rho^{'} \in \ball{\eps}{\rho}} \hmin{A}{B}_{\rho^{'}} .$$  	

\suppress{
\item {\bf Max-information}  For $\rho_{AB}\in \mathcal{D}(AB)$, $$ \imax(A:B)_{\rho} \defeq   \inf_{\sigma_{B}\in \mathcal{D}(B)}\dmax{\rho_{AB}}{\rho_{A}\otimes\sigma_{B}} .$$
\item {\bf Smooth max-information} For $\rho_{AB}\in \mathcal{D}(AB)$,  $$\imaxeps(A:B)_{\rho} \defeq \inf_{\rho'\in \ball{\eps}{\rho}} \imax(A:B)_{\rho'} .$$	

}
\end{enumerate}

We will use the following facts. 
\begin{fact}[Triangle inequality for purified distance,~\cite{Tomamichel12}]
\label{fact:trianglepurified}
For states $\rho_A, \sigma_A, \tau_A\in \mathcal{D}(A)$,
$$\Pur(\rho_A,\sigma_A) \leq \Pur(\rho_A,\tau_A)  + \Pur(\tau_A,\sigma_A) . $$ 
\end{fact}

\begin{fact}[ Fuchs-van de Graaf inequalities, \cite{FuchsG99}]
\label{fuchsG}
For states $\rho_A, \sigma_A\in \mathcal{D}(A)$,
$$\Delta(\rho_A, \sigma_A)\leq \Pur(\rho_A, \sigma_A) \leq \sqrt{2\Delta(\rho_A,\sigma_A)}.$$
\end{fact}

\suppress{\begin{fact}[\cite{stinespring55}](\textbf{Stinespring representation})\label{stinespring}
Let $\cE(\cdot): \mathcal{L}(A)\rightarrow \mathcal{L}(B)$ be a quantum operation. There exists a register $C$ and an unitary $U\in \mathcal{U}(ABC)$ such that $\cE(\omega)=\Tr_{A,C}\br{U (\omega  \otimes \ketbra{0}^{B,C}) U^{\dagger}}$. Stinespring representation for a channel is not unique. 
\end{fact}
}

\begin{fact}[Monotonicity under quantum operations, \cite{barnum96},\cite{lindblad75}]
	\label{fact:monotonequantumoperation}
For quantum states $\rho$, $\sigma \in \mathcal{D}(A)$, and quantum map $\cE(\cdot):\mathcal{L}(A)\rightarrow \mathcal{L}(B)$, it holds that
\begin{align*}
	\Pur(\cE(\rho),\cE(\sigma)) \leq \Pur(\rho,\sigma) \quad \mbox{and} \quad \F(\cE(\rho),\cE(\sigma)) \geq \F(\rho,\sigma) \quad \mbox{and} \quad \relent{\rho}{\sigma}\geq \relent{\cE(\rho)}{\cE(\sigma)}.
\end{align*}
\end{fact}

\begin{fact}[Uhlmann's Theorem, \cite{uhlmann76}]
\label{uhlmann}
Let $\rho_A,\sigma_A\in \mathcal{D}(A)$. Let $\rho_{AB}\in \mathcal{D}(AB)$ be a purification of $\rho_A$ and $\sigma_{AC}\in\mathcal{D}(AC)$ be a purification of $\sigma_A$. There exists an isometry $V: C \rightarrow B$ such that,
 $$\F(\ketbra{\theta}_{AB}, \ketbra{\rho}_{AB}) = \F(\rho_A,\sigma_A) ,$$
 where $\ket{\theta}_{AB} = (I_A \otimes V) \ket{\sigma}_{AC}$.
\end{fact}

\begin{fact}[Gentle measurement lemma,\cite{Winter:1999,Ogawa:2002}]
\label{gentlelemma}
Let $\rho$ be a quantum state and $0<A<I$ be an operator. Then 
$$\F(\rho, \frac{A\rho A}{\Tr(A^2\rho)})\geq \sqrt{\Tr(A^2\rho)}.$$
\end{fact}

\begin{fact}[Hayashi-Nagaoka inequality, \cite{hayashinagaoka} ]
\label{haynag}
Let $0<S<\id,T$ be positive semi-definite operators. Then 
$$\id-(S+T)^{-\frac{1}{2}}S(S+T)^{-\frac{1}{2}}\leq 2(\id-S) + 4T.$$
\end{fact}

\begin{fact}[Pinsker's inequality and a stronger statement, \cite{Jain:2003a}, \cite{CastelleHR78}]
\label{pinsker}
For quantum states $\rho_A,\sigma_A\in\mathcal{D}(A)$, 
$$\F(\rho_A,\sigma_A) \geq 2^{-\frac{1}{2}\relent{\rho_A}{\sigma_A}}.$$ This implies
$$\|\rho_A-\sigma_A\|_1^2 \leq 2\relent{\rho}{\sigma}.$$
\end{fact}

\begin{fact}[\cite{TomHay13, li2014}]
\label{dmaxequi}
Let $\eps\in (0,1)$ and $n$ be an integer. Let $\rho^{\otimes n}, \sigma^{\otimes n}$ be quantum states. Define $\Phi(x) = \int_{-\infty}^x \frac{e^{-t^2/2}}{\sqrt{2\pi}} dt$. It holds that
\begin{equation*}
\dmaxeps{\rho^{\otimes n}}{\sigma^{\otimes n}}{\eps} = n\relent{\rho}{\sigma} + \sqrt{n\varrelent{\rho}{\sigma}} \Phi^{-1}(\eps) + O(\log n) ,
\end{equation*}
\begin{equation*}
\dheps{\rho^{\otimes n}}{\sigma^{\otimes n}}{\eps^2} = n\relent{\rho}{\sigma} + \sqrt{n\varrelent{\rho}{\sigma}} \Phi^{-1}(\eps) + O(\log n).
\end{equation*}
\end{fact}

\begin{fact}
\label{gaussianupper}
For the function $\Phi(x) = \int_{-\infty}^x \frac{e^{-t^2/2}}{\sqrt{2\pi}} dt$ and $\eps\leq \frac{1}{2}$, it holds that $|\Phi^{-1}(\eps)| \leq 2\sqrt{\log\frac{1}{2\eps}}$.
\end{fact}
\begin{proof}
We have $$\Phi(-x)=\int_{-\infty}^{-x} \frac{e^{-t^2/2}}{\sqrt{2\pi}} dt = \int_{0}^{\infty} \frac{e^{-(-x-t)^2/2}}{\sqrt{2\pi}} dt \leq e^{-x^2/2} \int_{0}^{\infty} \frac{e^{-(-t)^2/2}}{\sqrt{2\pi}} dt = \frac{1}{2}e^{-x^2/2}.$$ Thus, $\Phi^{-1}(\eps) \geq -2\sqrt{\log\frac{1}{2\eps}}$, which completes the proof.
\end{proof}

\begin{fact}
\label{slowchange}
Let $\rho_1$ be a quantum state and $\{\cE_2,\cE_3, \ldots\}$ be a collection of quantum maps. Define a series of quantum states $\{\rho_2,\rho_3,\ldots \}$ recursively as $\rho_i = \cE_i(\rho_{i-1})$. It holds that $$\Pur(\rho_i,\rho_1) \leq (i-1)\max_i\{\Pur(\cE_i(\rho_1),\rho_1)\}.$$
\end{fact}
\begin{proof}
Consider 
$$\Pur(\rho_i, \rho_1) = \Pur(\cE_i(\rho_{i-1}),\rho_1) \leq \Pur(\cE_i(\rho_{i-1}),\cE_i(\rho_1)) + \Pur(\cE_i(\rho_1),\rho_1) \leq \Pur(\rho_{i-1},\rho_1) + \Pur(\cE_i(\rho_1),\rho_1).$$ This completes the proof by induction.
\end{proof}

\begin{fact}
\label{dmaxwithid}
Let $\rho_{AB}$ be a classical-quantum state with $B$ being the classical register. For every $\eps\in (0,1)$, it holds that
$$\dmaxeps{\rho_{AB}}{\rho_A\otimes \frac{\id_B}{|B|}}{\eps} \leq \log|B|.$$
\end{fact}
\begin{proof}
For some distribution $p(b)$, we have 
$$\rho_{AB} = \sum_b p(b)\rho^b_A\otimes \ketbra{b}_B \preceq \sum_b p(b)\rho^b_A\otimes \id_B = |B| \rho_A \otimes \frac{\id_B}{|B|}.$$ This implies that
$$\dmax{\rho_{AB}}{\rho_A\otimes \frac{\id_B}{|B|}} \leq \log|B|.$$
Since $\dmaxeps{\rho_{AB}}{\rho_A\otimes \frac{\id_B}{|B|}}{\eps} \leq \dmax{\rho_{AB}}{\rho_A\otimes \frac{\id_B}{|B|}}$ for all $\eps \in (0,1)$, the proof concludes.
\end{proof}

Following fact was implicitly present in \cite{Renner11} (see also Lemma 1 in \cite{AnshuJW17}). 

\begin{fact}
\label{imaxvariants}
For a quantum states $\rho_{AB}, \sigma_A, \sigma_B$, it holds that
$$\inf_{\rho'\in \ball{2\eps}{\rho_{AB}}}\dmax{\rho'_{AB}}{\rho'_A\otimes\sigma_B} \leq \dmaxeps{\rho_{AB}}{\sigma_A\otimes\sigma_B}{\eps} + 3\log\frac{2}{\eps}.$$
\end{fact}
\begin{proof}
Let $\rho''_{AB}\in \ball{\eps}{\rho_{AB}}$ be quantum state that achieves the optimum in $\dmaxeps{\rho_{AB}}{\sigma_A\otimes\sigma_B}{\eps}$. From \cite[Claim 5]{AnshuJW17} (a formal statement of an argument originally given in \cite{Renner11}), there exists a quantum state $\rho'_{AB}\in \ball{\eps}{\rho''_{AB}}$ such that 
$$\dmax{\rho'_{AB}}{\rho'_A\otimes \sigma_B} \leq \dmax{\rho''_{AB}}{\sigma_A\otimes \sigma_B} + 3\log\frac{2}{\eps}.$$
Since $\rho'_{AB}\in \ball{2\eps}{\rho_{AB}}$, the proof concludes.
\end{proof}

Following fact is about explicit constructions of pairwise independent hash functions.
\begin{fact}[Section 3, \cite{Lovettnotes}]
\label{pairwise}
Let $p$ be a prime, $k$ be a positive integer and $\cA, \cA'$, $\cB, \cB'$ be sets of size $|\cA|=|\cA'|=p^k$ and $|\cB|=|\cB'|=p$. There exist distinct functions $\{h_{a,b}: \cA'\rightarrow \cB', a\in \cA, b\in \cB\}$ such that for all $b'_1, b'_2\in \cB', a'_1,a'_2\in \cA'$
$$\frac{1}{p^{k+1}}|\{a,b: h_{a,b}(a'_1)=b'_1 \quad\&\quad  h_{a,b}(a'_2)=b'_2\}| = \frac{1}{p^2}.$$
\end{fact}

It leads to the following construction of pairwise independent probability distributions.
\begin{claim}
\label{pairwisedist}
Let $\cC$ be a set such that $|\cC|$ is a prime number and $n> 1$ be an integer. There exists a pairwise independent probability distribution $\barq(c_1,c_2, \ldots c_n)$, taking values over the set $\cC\times \cC \times \ldots \cC$ ($n$ times) such that $\barq(c_i) = \frac{1}{|\cC|}$ for all $i\in \{1,2, \ldots n\}$ and $c_i\in \cC$. The support size of $\barq$ is 
$|\cC|^{k+1}$, where $k \defeq \left\lceil \frac{\log n}{\log |\cC|}\right\rceil$.  Further, for any $i \leq n$ and $c_i \in \cC$, the distribution $\barq(c_1, \ldots c_{i-1}, c_{i+1}, \ldots c_n | c_i)$ is uniform in a support of size $|\cC|^{k}$.
\end{claim}
\begin{proof}
By definition, $k$ is the smallest integer such that $n\leq |\cC|^k$.  Let $\cA, \cB$ be two sets such that $|\cA| = |\cC|^k$ and $|\cB| = |\cC|$. Let $\cA' \defeq \{1,2, \ldots |\cC|^k\}$ and $\cB'=\cC$. Let $h_{a,b}: \cA' \rightarrow \cB'$ be the function obtained from Fact \ref{pairwise} for $p=|\cC|$. Define a distribution $\barq$ over $\cA \times \cB \times \cC^n$, where $\cC^n \defeq \cC\times \ldots \times \cC$ ($n$ times), as follows.
$$\barq(a,b) = \frac{1}{|\cC|^{k+1}}, \quad \barq(c_1, c_2, \ldots c_n \mid a,b) = 1 \mbox{ iff } h_{a,b}(i) = c_i \quad \forall i \leq n.$$
 Since $\{h_{a,b}\}$ are distinct functions, 
$$\barq(c_1, c_2, \ldots c_n) = \frac{1}{|\cC|^{k+1}} \mbox{ if } \exists (a,b): h_{a,b}(i) = c_i \quad \forall i \leq n,$$
and $0$ otherwise. Thus, the support size of $\barq$ over $\cC^n$ is equal to $|\cC|^{k+1}$. Using Fact \ref{pairwise}, we conclude that for all $i\neq j$ such that $i \leq n, j \leq n$, 
$$\barq(c_i, c_j) = \sum_{a, b} \barq(a,b)\barq(c_1, c_j\mid a,b) = \frac{|\{a, b: h_{a,b}(i)=c_i \quad\&\quad h_{a,b}(j)=c_j\}|}{|\cC|^{k+1}} = \frac{1}{|\cC|^2}.$$
Thus, $\barq$ is pairwise independent over the set $\cC^n$. Finally, fix an $i\leq n$, $c_i\in \cC$. Consider 
\begin{eqnarray*}
&&\barq(c_1, c_2, \ldots c_{i-1}, c_{i+1}, \ldots  c_n | c_i) = \frac{1}{\barq(c_i)}\cdot \barq(c_1, c_2, \ldots c_n)\\
&&= \frac{1}{|\cC|^{k}}\mbox{ if } \exists (a,b): h_{a,b}(i) = c_i \quad \forall i \leq n,
\end{eqnarray*}
Again using the fact that $\{h_{a,b}\}$ are distinct functions, we conclude that $\barq(c_1, c_2, \ldots c_{i-1}, c_{i+1}, \ldots  c_n | c_i)$ is uniformly distributed in a support of size $|\cC|^{k}$. This completes the proof.
\end{proof}

Given above claim, it is easy to construct a pairwise independent random variable $\barq$ with the marginal distribution $\barq(c_i)$ equal to a given distribution $q(c_i)$, assuming that $q$ takes rational values over $\cC$. For this, introduce a sufficiently large set $\cC'$ and consider a function $F: \cC' \rightarrow \cC$ which takes a probability distribution $p$ over $\cC'$ to a probability distribution $F(p)$ over $\cC$. The set $\cC'$ is chosen such that there exists a function $F$ that takes the uniform distribution over $\cC'$ to the distribution $q$. If the distribution $\barq$ over $\cC'\times \cC' \times \ldots \cC'$ is pairwise independent, then the distribution $F \times F \times \ldots F(\barq)$ over $\cC\times \cC \times \ldots \cC$ is also pairwise independent (as $F\times F(\barq(c_i, c_j)) = F(\barq(c_i))\cdot F(\barq(c_j))$). Thus, we obtain the desired construction for a large family of marginal distributions $q$. 

\section{A convex-split lemma with limited randomness}
\label{sec:convsplit}

The work \cite{AnshuDJ14} showed the following statement.

\begin{fact}[Convex-split lemma, \cite{AnshuDJ14}]
\label{convexcombusual}
Let $\rho_{PQ}\in\mathcal{D}(PQ)$ and $\sigma_Q\in\mathcal{D}(Q)$ be quantum states such that $\text{supp}(\rho_Q)\subset\text{supp}(\sigma_Q)$.  Let $k \defeq \dmax{\rho_{PQ}}{\rho_P\otimes\sigma_Q}$. Define the following state
\begin{equation*}
\tau_{PQ_1Q_2\ldots Q_n} \defeq  \frac{1}{n}\sum_{j=1}^n \rho_{PQ_j}\otimes\sigma_{Q_1}\otimes \sigma_{Q_2}\ldots\otimes\sigma_{Q_{j-1}}\otimes\sigma_{Q_{j+1}}\ldots\otimes\sigma_{Q_n}
\end{equation*}
on $n+1$ registers $P,Q_1,Q_2,\ldots Q_n$, where $\rho_{PQ_j}$ and $\sigma_{Q_j}$ are copies of $\rho_{PQ}$ and $\sigma_Q$ (for all $j\leq n$), respectively. Then
 $$ \relent{\tau_{PQ_1Q_2\ldots Q_n}}{\rho_P \otimes \sigma_{Q_1}\otimes\sigma_{Q_2}\ldots \otimes \sigma_{Q_n}} \leq \log(1+ \frac{2^k}{n}).$$ 
\end{fact}

When the quantum state $\rho_{PQ}$ is classical-quantum with register $Q$ being classical in the basis of $\sigma_Q$, that is, $\rho_{PQ} = \sum_c p(c)\rho^{c}_P\otimes\ketbra{c}_Q$ and $\sigma_Q = \sum_c q(c)\ketbra{c}_Q$, convex-split lemma implies Equation \ref{eq:avgeq}, as shown in the following claim.

\begin{corollary}
\label{coveringlemconvsplit}
Let $\rho_{PQ}\in\mathcal{D}(PQ)$ be a quantum state such that 
$$\rho_{PQ} = \sum_c p(c)\rho^c_P\otimes\ketbra{c}_Q.$$
Above, $\{p(c)\}_c$ is a probability distribution.
Then
 $$\sum_{c_1, \ldots c_n}p(c_1)\ldots p(c_n)\relent{\frac{1}{n}(\rho^{c_1}_P +\ldots \rho^{c_n}_P)}{\rho_P} \leq \log(1+ \frac{2^k}{n}).$$ This implies
 $$\sum_{c_1, \ldots c_n}p(c_1)\ldots p(c_n)\|\frac{1}{n}(\rho^{c_1}_P +\ldots \rho^{c_n}_P) - \rho_P\|_1 \leq \sqrt{\frac{2^k}{n}}.$$
\end{corollary}
\begin{proof}
We apply Fact \ref{convexcombusual} with $\sigma_Q = \rho_Q$ and consider the classical-quantum state $\tau_{PQ_1Q_2\ldots Q_n}$. It holds that
$$\tau^{c_1, \ldots c_n}_P = \frac{1}{n}(\rho^{c_1}_P +\ldots \rho^{c_n}_P).$$
Thus, we conclude from Fact \ref{convexcombusual} that 
 $$\sum_{c_1, \ldots c_n}p(c_1)\ldots p(c_n)\relent{\frac{1}{n}(\rho^{c_1}_P +\ldots \rho^{c_n}_P)}{\rho_P} \leq \log(1+ \frac{2^k}{n}).$$
Applying Fact \ref{pinsker}, this implies that  
$$\sum_{c_1, \ldots c_n}p(c_1)\ldots p(c_n)\|\frac{1}{n}(\rho^{c_1}_P +\ldots \rho^{c_n}_P) - \rho_P\|_1^2 \leq \frac{2^k}{n},$$ and the convexity of the square function leads to 
$$\sum_{c_1, \ldots c_n}p(c_1)\ldots p(c_n)\|\frac{1}{n}(\rho^{c_1}_P +\ldots \rho^{c_n}_P) - \rho_P\|_1 \leq \sqrt{\frac{2^k}{n}}.$$
This completes the proof.
\end{proof}

In the following, we will considerably improve upon above result in terms of additional randomness required. The main motivation is the observation that the proof of the convex-split lemma, as given in \cite{AnshuDJ14}, only requires some max-relative entropy bounds on quantum states involved in pairs of registers.

\begin{lemma}
\label{convsplitpairwise}
Let $\rho_{PQ}\in\mathcal{D}(PQ)$ and $\sigma_Q\in\mathcal{D}(Q)$ be quantum states such that 
$$\rho_{PQ} = \sum_c p(c)\rho^c_P\otimes\ketbra{c}_Q, \quad \sigma_Q = \sum_c q(c)\ketbra{c}_Q$$
and $\text{supp}(\rho_Q)\subset\text{supp}(\sigma_Q)$. Above, $\{p(c)\}_c, \{q(c)\}_c$ are probability distributions. Let $$\bsigma_{Q_1Q_2\ldots Q_n}:= \sum_{c_1\ldots c_n}\bar{q}(c_1\ldots c_n)\ketbra{c_1\ldots c_n}$$ be a quantum state that satisfies  $\bsigma_{Q_j}=\sigma_Q$ for all $j \leq n$ and $\barq$ is a pairwise independent probability distribution.  Let $k \defeq \dmax{\rho_{PQ}}{\rho_P\otimes\sigma_Q}$. Define the following states
\begin{equation*}
\tau^{(j)}_{PQ_1Q_2\ldots Q_n}= \sum_cp(c)\rho^c_P\otimes \ketbra{c}_{Q_j}\otimes \bsigma^{c}_{Q_1\ldots Q_{j-1}Q_{j+1}\ldots Q_n}, \quad \tau_{PQ_1Q_2\ldots Q_n} \defeq  \frac{1}{n}\sum_{j=1}^n \tau^{(j)}_{PQ_1Q_2\ldots Q_n}
\end{equation*}
on $n+1$ registers $P,Q_1,Q_2,\ldots Q_n$. Then 
 $$ \relent{\tau_{PQ_1Q_2\ldots Q_n}}{\rho_P \otimes \bsigma_{Q_1\ldots Q_n}} \leq \log (1+\frac{2^k}{n}).$$ 
\end{lemma}
\begin{proof}
Consider the following identity (as shown in \cite[Supplementary Material]{AnshuDJ14}) which can be verified by direct calculation.
\begin{eqnarray}
\label{eq:convsplit}
 \relent{\tau_{PQ_1\ldots Q_n}}{\rho_P\otimes \bsigma_{Q_1\ldots Q_n}} &=& \frac{1}{n}\sum_j \relent{\tau^{(j)}_{PQ_1Q_2\ldots Q_n}}{\rho_P\otimes \bsigma_{Q_1\ldots Q_n}} \nonumber\\ &-& \frac{1}{n}\sum_j\relent{\tau^{(j)}_{PQ_1Q_2\ldots Q_n}}{\tau_{PQ_1\ldots Q_n}}.
\end{eqnarray}
Define the map $\cR_j: Q_j\rightarrow Q_1Q_2\ldots Q_n$ as $\cR_j(\ketbra{c}_{Q_j}) = \ketbra{c}_{Q_j}\otimes \bsigma^c_{Q_1\ldots Q_{j-1}Q_{j+1}\ldots Q_n}$. Then 
\begin{eqnarray*}
\relent{\tau^{(j)}_{PQ_j}}{\rho_P\otimes \sigma_{Q_j}}&\leq&\relent{\tau^{(j)}_{PQ_1Q_2\ldots Q_n}}{\rho_P\otimes \bsigma_{Q_1\ldots Q_n}} \\ &=& \relent{\id_P\otimes\cR_j(\tau^{(j)}_{PQ_j})}{\id_P\otimes\cR_j(\rho_P\otimes \sigma_{Q_j})}\\ &\leq&\relent{\tau^{(j)}_{PQ_j}}{\rho_P\otimes \sigma_{Q_j}}.
\end{eqnarray*}
Since $\tau^{(j)}_{PQ_j} = \rho_{PQ_j}$, we obtain $\relent{\tau^{(j)}_{PQ_1Q_2\ldots Q_n}}{\rho_P\otimes \bsigma_{Q_1\ldots Q_n}} = \relent{\rho_{PQ_j}}{\rho_P\otimes \sigma_{Q_j}}$.

The second term $\relent{\tau^{(j)}_{PQ_1Q_2\ldots Q_n}}{\tau_{PQ_1\ldots Q_n}}$ is lower bounded by $\relent{\tau^{(j)}_{PQ_j}}{\tau_{PQ_j}} = \relent{\rho_{PQ_j}}{\tau_{PQ_j}}$. But observe that 
\begin{eqnarray*}
\tau_{PQ_j} &=& \frac{1}{n}\tau^{(j)}_{PQ_j}+ \frac{1}{n}\sum_{j'\neq j}\tau^{(j')}_{PQ_{j}}\\ 
&=& \frac{1}{n}\rho_{PQ_j}+ \frac{1}{n}\sum_{j'\neq j}\sum_cp(c)\rho^c_P\otimes \bsigma^c_{Q_{j}}\\
&=& \frac{1}{n}\rho_{PQ_j}+ \frac{1}{n}\sum_{j'\neq j}\sum_cp(c)\rho^c_P\otimes \sigma_{Q_{j}}\\
&=& \frac{1}{n}\rho_{PQ_j}+ \frac{n-1}{n}\rho_P\otimes \sigma_{Q_{j}}
\end{eqnarray*}
where second last equality follows since $\bsigma_{Q_jQ_{j'}}=\sigma_{Q_j}\otimes \sigma_{Q_{j'}}$ (as $\barq$ is pairwise independent).

By assumption, $\rho_{PQ_j} \leq 2^k\rho_P\otimes \sigma_{Q_j}$. Hence $\tau_{PQ_j} \leq (1+\frac{2^k-1}{n})\rho_P\otimes\sigma_{Q_j}$. Since $\log(A)\leq \log(B)$ if $A \leq B$ for positive semidefinite matrices $A$ and $B$ (see for example, \cite{carlen}), we have
\begin{eqnarray*}
&&\relent{\rho_{PQ_j}}{\tau_{PQ_j}} = \Tr(\rho_{PQ_j}\log\rho_{PQ_j}) - \Tr(\rho_{PQ_j}\log\tau_{PQ_j}) \\ &&\geq \Tr(\rho_{PQ_j}\log\rho_{PQ_j})  - \Tr(\rho_{PQ_j}\log(\rho_P\otimes\sigma_{Q_j})) - \log(1 + \frac{2^k-1}{n}) \\ && = \relent{\rho_{PQ_j}}{\rho_P\otimes\sigma_{Q_j}}- \log(1 + \frac{2^k-1}{n}) .
\end{eqnarray*}
Using in Equation \ref{eq:convsplit}, we find that
\begin{eqnarray*}
\relent{\tau_{PQ_1\ldots Q_n}}{\rho_P\otimes \bsigma_{Q_1\ldots Q_n}}&\leq& \frac{1}{n}\sum_j \relent{\rho_{PQ_j}}{\rho_P\otimes\sigma_{Q_j}} - \frac{1}{n}\sum_j\relent{\rho_{PQ_j}}{\rho_P\otimes\sigma_{Q_j}}+ \log(1 + \frac{2^k-1}{n})\\ &=& \log(1 + \frac{2^k-1}{n}).
\end{eqnarray*}
Thus, the lemma follows.  

\end{proof}

A corollary of Lemma \ref{convsplitpairwise} is as follows.
\begin{corollary}
\label{corconvex}
Fix an $\eps\in (0,1)$. Let $\rho_{PQ}\in\mathcal{D}(PQ)$ and $\sigma_Q\in\mathcal{D}(Q)$ be quantum states such that 
$$\rho_{PQ} = \sum_c p(c)\rho^c_P\otimes\ketbra{c}_Q, \quad \sigma_Q = \sum_c q(c)\ketbra{c}_Q$$
and $\text{supp}(\rho_Q)\subset\text{supp}(\sigma_Q)$. Above, $\{p(c)\}_c, \{q(c)\}_c$ are probability distributions. Let $$\bsigma_{Q_1Q_2\ldots Q_n}:= \sum_{c_1\ldots c_n}\bar{q}(c_1\ldots c_n)\ketbra{c_1\ldots c_n}$$ be a quantum state that satisfies  $\bsigma_{Q_j}=\sigma_Q$ for all $j\leq n$ and $\barq$ is a pairwise independent probability distribution.  Let $k \defeq \min_{\rho'_{PQ}\in \ball{\eps}{\rho_{PQ}}}\dmax{\rho'_{PQ}}{\rho'_P\otimes \sigma_Q}$. Define the following states
\begin{equation*}
\tau^{(j)}_{PQ_1Q_2\ldots Q_n}= \sum_cp(c)\rho^c_P\otimes \ketbra{c}_{Q_j}\otimes \bsigma^{c}_{Q_1\ldots Q_{j-1}Q_{j+1}\ldots Q_n}, \quad \tau_{PQ_1Q_2\ldots Q_n} \defeq  \frac{1}{n}\sum_{j=1}^n \tau^{(j)}_{PQ_1Q_2\ldots Q_n}
\end{equation*}
on $n+1$ registers $P,Q_1,Q_2,\ldots Q_n$, where $\forall j \in [n]: \rho_{PQ_j} = \rho_{PQ}$. For $\delta\in (0,1)$ and $n\geq \lceil \frac{2^k}{\delta^2}\rceil$,  
 $$ \Pur(\tau_{PQ_1Q_2\ldots Q_n}, \rho_P \otimes \bsigma_{Q_1\ldots Q_n}) \leq 2\eps + \delta.$$ 
\end{corollary}
\begin{proof}
Let $\rho'_{PQ}$ achieve the minimum in the definition of $k$. Since the quantum state $\rho''_{PQ}$, obtained by measuring $\rho'_{PQ}$ in the eigenbasis of $\sigma_Q$, satisfies $\rho''_{PQ}\in \ball{\eps}{\rho_{PQ}}$ and $\dmax{\rho''_{PQ}}{\rho''_P\otimes \sigma_Q}\leq k$ (by Fact \ref{fact:monotonequantumoperation}), we can assume that $\rho'_{PQ}$ itself is a classical quantum state. 

We apply Lemma \ref{convsplitpairwise} to quantum states $\rho'_{PQ}, \sigma_Q$ to obtain that 
$$\relent{\tau'_{PQ_1Q_2\ldots Q_n}}{\rho'_P \otimes \bsigma_{Q_1\ldots Q_n}} \leq \log (1+\frac{2^{k}}{n})\leq \log(1+\delta^2),$$ where 
\begin{equation*}
\tau'_{PQ_1Q_2\ldots Q_n} \defeq  \frac{1}{n}\sum_{j=1}^n \tau'^{(j)}_{PQ_1Q_2\ldots Q_n},\quad \tau'^{(j)}_{PQ_1Q_2\ldots Q_n}= \sum_cp(c)\rho'^c_P\otimes \ketbra{c}_{Q_j}\otimes \bsigma^{c}_{Q_1\ldots Q_{j-1}Q_{j+1}\ldots Q_n}. 
\end{equation*}
Observe that $\Pur(\tau'_{PQ_1Q_2\ldots Q_n}, \tau_{PQ_1Q_2\ldots Q_n})\leq \eps$.

From Pinsker's inequality (Fact \ref{pinsker}) we conclude that $\F^2(\tau'_{PQ_1Q_2\ldots Q_n}, \rho'_P \otimes \bsigma_{Q_1\ldots Q_n}) \geq \frac{1}{1+\delta^2}$, or $\Pur(\tau'_{PQ_1Q_2\ldots Q_n}, \rho'_P \otimes \bsigma_{Q_1\ldots Q_n}) \leq \delta$. The lemma concludes by triangle inequality for purified distance. 
\end{proof}

\section{An achievability bound for quantum measurement compression}
\label{sec:quantumstate}

We formally introduce the task, starting from the point where Alice has already `coherently' applied the a quantum measurement on her registers. 
\begin{definition}[Quantum measurement compression]
\label{def:qmc}
Fix an $\eps\in (0,1)$, and consider the state $\ketbra{\Psi}_{RAC\bar{C}B}$ of the form $\sum_c \sqrt{p(c)} \ket{c}_C \ket{c}_{\bar{C}} \ket{\psi^c}_{RAB}$. An $(m, r_1, r_2, \eps)$- quantum measurement compression protocol consists of 
\begin{itemize}
\item a classical-classical state (or preshared randomness) $\theta_{E_AE_B}$ between Alice ($E_A$) and Bob ($E_B$), 
\item an encoding isometry by Alice $V: \cH_{AC\bar{C}E_A} \rightarrow \cH_{AC\bar{C}QT_AT'_A}$, and
\item a decoding isometry by Bob $W: \cH_{QBE_B}\rightarrow \cH_{BC'T_BT'_B}$, where $C'\equiv C$
\end{itemize}
such that 
$$\Pur\left(\Tr_{T'_AT'_B\bar{C}}\left(WV(\Psi_{RABC}\otimes \theta_{E_AE_B})V^{\dagger}W^{\dagger}\right),\Psi_{RABCC'}\otimes \omega_{T_A T_B}\right)\leq \eps,$$ for some classical-classical state $\omega_{T_AT_B}$.
The number of bits communicated is $m=\log|Q|$, number of bits of initial shared randomness is $r_1= \max(\log|E_A|, \log|E_B|)$ and final number of bits of shared randomness is $r_2 = \min(\log|T_A|, \log|T_B|)$. 
\end{definition}
Note that in the above definition, the registers $T'_A, T'_B$ may be arbitrarily correlated with other registers, but they are discarded and do not count in the randomness gained. The registers $T_A, T_B$ contain the gained randomness. Note that the gained randomness is independent of $\Psi_{RABCC'}$, which is important since this randomness can be used for future tasks.

We prove the following theorem for the task in Definition \ref{def:qmc}. The randomness consumed below is characterized by $\dmaxeps{\Psi_{RBC}}{\Psi_{RB}\otimes \id_C}{\eps}$, which is closely related to $\hmineps{C}{RB}{\eps}$, except that the optimization over register $RB$ is not present.
\begin{theorem}
\label{newcompression}
Let $\ketbra{\Psi}_{RAC\bar{C}B}\in \cD(RAC\bar{C}B)$ be a pure quantum state of the form $\sum_c \sqrt{p(c)} \ket{c}_C \ket{c}_{\bar{C}} \ket{\psi^c}_{RAB}$, $\sigma_C\in \cD(C)$ be a quantum state satisfying $\supp(\Psi_C)\subseteq \supp(\sigma_C)$ that commutes with $\Psi_C$ and $\eps \in (0,1/10)$. There exists a $(m, r_1, r_2, 10\eps)$- quantum measurement compression protocol with 
$$m\leq \dmaxeps{\Psi_{RBC}}{\Psi_{RB}\otimes \sigma_C}{\eps} - \dheps{\Psi_{BC}}{\Psi_{B}\otimes \sigma_C}{\eps^2} + 7\log\left(\frac{1}{\varepsilon}\right).$$ If $\sigma_C$ is chosen to be the uniform distribution, then
$$ r_1\leq 2\log|C| + \log\frac{8}{\eps^5}$$ and
 $$r_2 \geq 2\log|C| + \dmaxeps{\Psi_{RBC}}{\Psi_{RB}\otimes \id_C}{\eps} +  \log\frac{8}{\eps^5} .$$
\end{theorem}

\noindent {\it Outline of the proof:} In the protocol, Alice and Bob pre-share the pairwise independent randomness in $n$ registers (where $n$ is to be specified below), where the registers are divided into several blocks of size $b$ each ($b$ is again to be specified below). They start with the quantum state where the measurement has already been performed (that is, the quantum state $\ket{\Psi}_{RAC\bar{C}B}$) . This step in itself is a point of departure from the proof given in \cite{Winter04, WildeHBH12}, where the POVM was decomposed into a convex combination of other POVMs. On the other hand, in our protocol, Alice only focuses the post measurement state. Conditioned on a sample from the shared randomness, Alice applies an appropriate unitary on her registers which correlates the register $C$ with a location on the shared randomness. If Alice had communicated this location to Bob, the task would be completed as Bob would be able to pick up the correct randomness. Instead, she only tells the block number to Bob, who then finds out the correct location by performing quantum hypothesis testing.  

\begin{proof}

We begin with defining some important quantities for the proof.
\vspace{0.1in}

\noindent {\bf 1. Quantum states and registers involved in the proof:} Let $\sigma_C$ be of the form $\sigma_C = \sum_c q(c)\ketbra{c}_C$. Let 
$$k \defeq \dmaxeps{\Psi_{RBC}}{\Psi_{RB}\otimes \sigma_C}{\eps}, \quad n \defeq \lceil 8\cdot\frac{2^k }{\eps^5}\rceil, \quad b\defeq \lceil \eps^2\cdot 2^{\dheps{\Psi_{BC}}{\Psi_{B}\otimes \sigma_C}{\eps^2}}\rceil.$$ Let 
$$k' \defeq \min_{\Psi'_{RBC}\in \ball{2\eps}{\Psi_{RBC}}}\dmax{\Psi'_{RBC}}{\Psi'_{RB}\otimes \sigma_C}.$$
From Fact \ref{imaxvariants}, it holds that $k' \leq k + \log\frac{8}{\eps^3}$. Thus, $n \geq \lceil\frac{2^{k'}}{\eps^2}\rceil$.
The assumption that $\supp(\Psi_C)\subseteq \supp(\sigma_C)$ ensures that all the above quantities are well defined. By definition of $\dheps{\Psi_{BC}}{\Psi_{B}\otimes \sigma_C}{\eps^2}$, there exists a projector $\Pi_{BC}$ such that $$\Tr(\Pi_{BC}\Psi_{BC})\geq 1-\eps^2, \quad \Tr(\Pi_{BC}\Psi_{B}\otimes \sigma_C)\leq \eps^2/b.$$ Let $\sigma_{CC'} \defeq \sum_c q(c)\ketbra{c}_{C}\otimes\ketbra{c}_{C'}$ be an extension of $\sigma_C$ and $$\bsigma_{C_1C'_1\ldots C_nC'_n}:= \sum_{c_1,\ldots c_n}\bar{q}(c_1,\ldots c_n)\ketbra{c_1,\ldots c_n}_{C_1\ldots C_n}\otimes \ketbra{c_1,\ldots c_n}_{C'_1\ldots C'_n}$$ be the quantum state such that the probability distribution $\{\bar{q}(c_1,\ldots c_n)\}_{c_1,\ldots c_n}$ is pairwise independent and satisfies $\bar{q}(c_i) = q(c_i)$. 
Define the quantum states,
$$\xi_{RB C_1C'_1\ldots C_nC'_n}  \defeq \Psi_{RB}\otimes \bsigma_{C_1C'_1\ldots C_nC'_n},$$
$$\xi_{RABC\bar{C} C_1\ldots C_nC'_1\ldots C'_n} \defeq   \ketbra{\Psi}_{RABC\bar{C}} \otimes  \bsigma_{C_1C'_1\ldots C_nC'_n},$$ 
and
$$\mu_{RB C_1C'_1 \ldots C_nC'_n} \defeq \frac{1}{n}\sum_{j=1}^n\sum_c p(c) \psi^c_{RB}\otimes\ketbra{c, c}_{C_jC'_j}\otimes \bsigma^c_{C_1C'_1\ldots C_{j-1}C'_{j-1}C_{j+1}C'_{j+1}\ldots C_nC'_n}.$$
Note that $\Psi_{RB} = \mu_{RB}$. Consider,
\begin{eqnarray*}
&&\mu_{RB C_1C'_1 \ldots C_nC'_n} = \frac{1}{n}\sum_{j=1}^n\sum_c p(c) \psi^c_{RB}\otimes\ketbra{c, c}_{C_jC'_j}\otimes \bsigma^c_{C_1C'_1\ldots C_{j-1}C'_{j-1}C_{j+1}C'_{j+1}\ldots C_nC'_n}\\
&& \overset{a}= \frac{1}{n}\sum_{j=1}^n\sum_{c_j} p(c_j) \psi^{c_j}_{RB}\otimes\ketbra{c_j, c_j}_{C_jC'_j}\otimes \bsigma^{c_j}_{C_1C'_1\ldots C_{j-1}C'_{j-1}C_{j+1}C'_{j+1}\ldots C_nC'_n}\\
&& \overset{b}=\frac{1}{n}\sum_{j=1}^n\sum_{c_1, \ldots c_n} p(c_j) \psi^{c_j}_{RB}\otimes\barq(c_1, \ldots c_{j-1}c_{j+1}\ldots c_n\mid c_j)\ketbra{c_1, c_1\ldots c_{n},c_{n}}_{C_1C'_1\ldots C_nC'_n}\\
&& = \frac{1}{n}\sum_{j=1}^n\sum_{c_1, \ldots c_n} \frac{p(c_j)}{q(c_j)} \psi^{c_j}_{RB}\otimes\barq(c_1\ldots c_n)\ketbra{c_1, c_1\ldots c_{n},c_{n}}_{C_1C'_1\ldots C_nC'_n}\\
&& = \sum_{c_1\ldots c_n}\barq(c_1\ldots c_n)\ketbra{c_1\ldots c_n}_{C_1\ldots C_n}\otimes\ketbra{c_1\ldots c_n}_{C'_1\ldots C'_n}\otimes\left(\frac{1}{n}\sum_{j=1}^n\frac{p(c_j)}{q(c_j)} \psi^{c_j}_{RB} \right),
\end{eqnarray*}
where $(a)$ follows by renaming $c\rightarrow c_j$ and $(b)$ follows from the fact that $\sigma^{c_j}$ is the quantum state conditioned on the value $c_j$. Let $$\gamma(c_1,c_2,\ldots c_n) \defeq \frac{1}{n}\sum_{j=1}^n \frac{p(c_j)}{q(c_j)}$$ be a normalization parameter. Introducing a new register $J$, define the following quantum state for every $c_1,c_2\ldots c_n$: $$\ket{\theta^{c_1,c_2\ldots c_n}}_{JRAB} \defeq \frac{1}{\sqrt{\gamma(c_1,c_2,\ldots c_n)}}\sum_{j=1}^n \frac{1}{\sqrt{n}}\sqrt{\frac{p(c_j)}{q(c_j)}}\ket{\psi^{c_j}}_{RAB}\ket{j}_J.$$
Define an extension of $\mu_{RB C_1C'_1\ldots C_nC'_n}$ as, 
\begin{eqnarray*}
&&\mu_{RBAJC_1C'_1\ldots C_nC'_n} \defeq \\
&& \sum_{c_1\ldots c_n}\gamma(c_1,\ldots c_n)\cdot \barq(c_1\ldots c_n)\ketbra{c_1\ldots c_n}_{C_1\ldots C_n}\otimes\ketbra{c_1\ldots c_n}_{C'_1\ldots C'_n}\\ &&\otimes \ketbra{\theta^{c_1\ldots c_n}}_{JRAB}.
\end{eqnarray*}
Using the Corollary ~\ref{corconvex} and choice of $n$ we have,
$$\Pur(\xi_{RBC_1C'_1\ldots C_nC'_n}, \mu_{RB C_1C'_2\ldots C_nC'_n}) \leq 5\eps .$$ 
Thus, using Claim \ref{fact:cquhlmann}, we find that there exists an isometry depending on $c_1,c_2,\ldots c_n$: $U^{c_1\ldots c_n}: \cH_{AC\bar{C}}\rightarrow \cH_{AJ}$   
such that,
\begin{eqnarray}
\label{eq:conditionaUclose}
&&\Pur\bigg(\bigg(\sum_{c_1\ldots c_n}\ketbra{c_1\ldots c_n}\otimes U^{c_1\ldots c_n}\bigg) \xi_{RAC\bar{C}B C_1C'_1\ldots C_nC'_n}\nonumber\\ &&\bigg(\sum_{c_1\ldots c_n}\ketbra{c_1\ldots c_n}\otimes U^{c_1\ldots c_n\dagger}\bigg), \mu_{RBAJ C_1C'_1\ldots C_nC'_n}\bigg)  \nonumber\\ &&= \Pur(\xi_{RBC_1C'_1\ldots C_nC'_n}, \mu_{RB C_1C'_1\ldots C_nC'_n})  \leq 5\eps .
\end{eqnarray}

\vspace{0.1in}

\noindent {\bf 2. The protocol:} Consider the following protocol $\cP$.
\begin{enumerate}
\item Alice and Bob share the quantum state $\bsigma_{C_1C'_1\ldots C_nC'_n}$, where Alice holds the registers $C_1, C_2, \ldots C_n$ and Bob holds the registers $C'_1, C'_2, \ldots C'_n$. Alice, Bob and Reference share the quantum state $\ket{\Psi}_{RABC\bar{C}}$ between themselves where Alice holds the registers $AC\bar{C}$, Reference holds the register $R$ and Bob holds the registers $B$. 
\item Conditioned on the values $c_1,c_2\ldots c_n$ in registers $C_1C_2\ldots C_n$, Alice applies the isometry $U^{c_1,c_2\ldots c_n}$ on her register $AC\bar{C}$.
\begin{itemize}
\item The resulting state is close to the quantum state $\mu_{RBAJC_1C'_1\ldots C_nC'_n}$, by Equation \ref{eq:conditionaUclose}.
\end{itemize}
\item Alice measures the register $J$ and obtains the measurement outcome $j \in [n]$. She sends the integer $j_1\defeq \lfloor (j-1)/b \rfloor$ to Bob in a register $J_1$ using classical communication. Let $j_2 \defeq  j \hspace{1mm} \% \hspace{1mm} b$ be held by Alice in a register $J_2$. Here, $j\hspace{1mm} \% \hspace{1mm} b$ is equal to $j \text{ mod } b$ (where $\text{mod}$ is the remainder function) if $j<b$, and is equal to $b$ otherwise. 

\item Bob swaps the registers $C'_{b\cdot j_1+1},C'_{b\cdot j_1+2},\ldots C'_{b\cdot j_1+b}$ with the set of registers $C'_1,C'_2,\ldots C'_b$ in that order. In the same fashion, Alice swaps the registers $C_{b\cdot j_1+1},C_{b\cdot j_1+2},\ldots C_{b\cdot j_1+b}$ with the set of registers $C_1,C_2,\ldots C_b$ in that order. 
\begin{itemize}
\item If the shared state in step $2$ were $\mu_{RBAJC_1C'_1\ldots C_nC'_n}$, the joint quantum state in the registers $RBAJ_2C'_1C_1\ldots C'_bC_b$ at this step would be 
\begin{eqnarray*}
&&\mu^{(2)}_{RBAJ_2C'_1C_1\ldots C'_bC_b} \\ &&= \sum_{c_1\ldots c_b}\barq(c_1\ldots c_b)\ketbra{c_1\ldots c_b}_{C_1\ldots C_b}\otimes\ketbra{c_1\ldots c_b}_{C'_1\ldots C'_b}\otimes \\ && \hspace{2mm}\left(\frac{1}{b}\sum_{j_2=1}^b\frac{p(c_{j_2})}{q(c_{j_2})}\ketbra{\psi^{c_{j_2}}}_{RAB}\otimes \ketbra{j_2}_{J_2}\right)\\
&& = \frac{1}{b}\sum_{j_2=1}^b\sum_c p(c) \psi^c_{RB}\otimes\ketbra{c,c}_{C_{j_2}C'_{j_2}}\otimes \\ && \hspace{4mm} \bsigma^c_{C_1C'_1\ldots C_{j_2-1}C'_{j_2-1}C_{j_2+1}C'_{j_2+1}\ldots C_bC'_b}\otimes \ketbra{j_2}_{J_2}.
\end{eqnarray*}
\end{itemize}
\item Define,
$$\Pi_{j'_2}\defeq  \id_{C'_1}\otimes\ldots \id_{C'_{j'_2-1}}\otimes \Pi_{BC'_{j'_2}}\otimes \id_{C'_{j'_2+1}}\ldots \otimes \id_{C'_b} \quad \mbox{and} \quad \Pi\defeq \sum_{j'_2=1}^b \Pi_{j'_2} .$$ Bob applies the measurement (the hypothesis testing measurement)
\begin{eqnarray*}
\cA(X) &=& \sum_{j'_2=1}^b \left(\sqrt{\Pi^{-\frac{1}{2}}\Pi_{j'_2}\Pi^{-\frac{1}{2}}}\right)X\left(\sqrt{\Pi^{-\frac{1}{2}}\Pi_{j'_2}\Pi^{-\frac{1}{2}}}\right) \otimes \ketbra{j'_2}_{J'_2}\\ &&+ \left(\sqrt{\id- \Pi^0}\right)X\left(\sqrt{\id - \Pi^0}\right) \otimes \ketbra{0}_{J'_2},
\end{eqnarray*}
 where $J'_2$ is the outcome register and $\Pi^0$ is the projector onto the support of $\Pi$. 
\item Upon obtaining the outcome $j'_2$ (if not equal to $0$), Bob swaps $C'_{j'_2},C'_1$. If the outcome is equal to $0$, Bob performs no operation. He computes $j' = j_1\cdot b + j'_2$ and stores the value in a register $J'$.
\item Alice swaps the registers $C_{j_2}, C_1$. We note that this step could also be performed after Step $3$.
\item Final quantum state is obtained in the registers $RABC_1C'_1C_2C'_2JJ'$, where the registers $RABC_1C'_1$ contain the actual output and the registers $C_2C'_2JJ'$ contain the returned shared randomness. We represent it as $\Phi'_{RABC_1C'_1C_2C'_2JJ'}$.
\begin{itemize}
\item If the shared state in step $2$ were $\mu_{RBAJC_1C'_1\ldots C_nC'_n}$, let the final quantum state in registers $RABC_1C'_1C_2C'_2JJ'$ be  $\Phi^1_{RABC_1C'_1C_2C'_2JJ'}$.
\end{itemize}
\end{enumerate}

Define 
$$\mu^{ideal}_{RABCC_1C'_1C_2C'_2JJ'} \defeq  \Psi_{RBAC_1C'_1}\otimes \left(\frac{1}{n}\sum_j \ketbra{j,j}_{JJ'}\right)\otimes \bsigma_{C_2C'_2}.$$
We have the following claim. 
\begin{claim}[Hypothesis testing succeeds well]
\label{claimprotp1meas}
 It holds that $\Pur(\Phi^1, \mu^{ideal}) \leq \sqrt{18}\eps$.
\end{claim}
\begin{proof}
Applying the measurement $\cA$ to the quantum state $\mu^{(2)}_{RABC_1C'_1\ldots C_bC'_b}$, we obtain the quantum state $\cA(\mu^{(2)}_{RABC_1C'_1\ldots C_bC'_b})$.  Let the conditional probabilities $p_{j'_2|j_2}$ be defined as follows:
$$p_{j'_2|j_2} \defeq \Tr\left(\Pi^{-\frac{1}{2}}\Pi_{j'_2}\Pi^{-\frac{1}{2}}\sum_c p(c) \psi^c_{RB}\otimes\ketbra{c}_{C'_{j_2}}\otimes \bsigma^c_{C'_1\ldots C'_{j_2-1}C'_{j_2+1}\ldots C'_b}\right),$$
$$p_{0|j_2} \defeq \Tr\left((\id - \Pi^0)\sum_c p(c) \psi^c_{RB}\otimes\ketbra{c}_{C'_{j_2}}\otimes \bsigma^c_{C'_1\ldots C'_{j_2-1}C'_{j_2+1}\ldots C'_b}\right).$$
Define the quantum state 
\begin{eqnarray*}
\mu^{(4)}&\defeq& \frac{1}{b}\sum_{j_2}\sum_c p(c) \psi^c_{RB}\otimes\ketbra{c,c}_{C'_{j_2}C_{j_2}}\otimes \bsigma^c_{C_1C'_1\ldots C_{j_2-1}C'_{j_2-1}C_{j_2+1}C'_{j_2+1}\ldots C_bC'_b}\\ && \hspace{1cm} \otimes \ketbra{j_2, j_2}_{J_2, J'_2}.
\end{eqnarray*}
From Claim \ref{gentlepovm}, we find that 
$$\F(\cA(\mu^{(2)}),\mu^{(4)}) \geq (\frac{1}{b}\sum_{j_2} p_{j_2|j_2})^{3/2}.$$
Now using Hayashi-Nagaoka inequality (Fact \ref{haynag}) with $c=1$, we obtain
\begin{eqnarray*}
\frac{1}{b}\sum_{j'_2\neq j_2}p_{j'_2| j_2} &\leq& \frac{2}{b}\sum_{j_2}\Tr\bigg((I-\Pi_{j_2})\sum_c p(c) \psi^c_{RB}\otimes\ketbra{c}_{C_{j_2}}\\ &&\otimes \bsigma^c_{C_1C'_1\ldots C_{j_2-1}C'_{j_2-1}C_{j_2+1}C'_{j_2+1}\ldots C_bC'_b}\bigg)\\ &+&\frac{4}{b}\sum_{j_2}\Tr\bigg((\sum_{j'_2\neq j_2}\Pi_{j'_2})\sum_c p(c) \psi^c_{RB}\otimes\ketbra{c}_{C_{j_2}}\\ &&\otimes \bsigma^c_{C_1C'_1\ldots C_{j_2-1}C'_{j_2-1}C_{j_2+1}C'_{j_2+1}\ldots C_bC'_b}\bigg)
\\ &=& \frac{2}{b}\sum_{j_2} \Tr((I-\Pi_{j_2})\Psi_{BC'_{j_2}}) + \frac{4}{b}\sum_{j_2} \Tr((\sum_{j'_2\neq j_2}\Pi_{j'_2})\Psi_{B}\otimes \sigma_{C'_{j_2}}) \\
&\leq& 2\eps^2 + 4(b-1)\frac{\eps^2}{b} \leq 6\eps^2.
\end{eqnarray*}
The equality above uses the pairwise independence of $\barq$. This implies that 
$$\frac{1}{b}\sum_i p_{i|i} = \frac{1}{b}\sum_{i,j_2}p_{i|j_2} - \frac{1}{b}\sum_{i\neq j_2}p_{i|j_2} \geq 1-6\eps^2.$$ 
Thus, 
$$\F^2(\mu^4,\cA(\mu^2)) \geq (1-6\eps^2)^3 \geq 1-18\eps^2. $$
Now, Bob swaps registers $C'_{j'_2}$ and $C'_1$, controlled on value $j'_2$ in register $J'_2$, and Alice swaps registers $C_{j_2},C_1$ controlled on the value $j_2$ in $J_2$. These operations on the quantum state $\mu^4$ give the quantum state $\mu^{ideal}_{RBAC_1C'_1C_2C'_2JJ'}$ in registers $RBAC_1C'_1C_2C'_2JJ'$, using the pairwise independence of $\barq$. The claim now follows by the monotonicity of purified distance under quantum operations (Fact \ref{fact:monotonequantumoperation}).
\end{proof}

\vspace{0.1in}

\noindent\textbf{3. Analysis of the protocol:} Since quantum maps (the entire protocol $\cP$ can be viewed as a quantum map from input to output) do not decrease fidelity (monotonicity of fidelity under quantum operation, Fact~\ref{fact:monotonequantumoperation}), we have,
\begin{equation}
\Pur(\Phi^1_{RABC_1C'_1C_2C'_2JJ'}, \Phi'_{RABC_1C'_1C_2C'_2JJ'}) \leq 5\eps.
\label{eq:prot}
\end{equation}
This implies using Claim \ref{claimprotp1meas} and triangle inequality for purified distance \ref{fact:trianglepurified} that 
\begin{equation}
\label{eq:finaldist}
\Pur(\Phi'_{RABC_1C'_1C_2C'_2JJ'},\mu^{ideal}_{RABC_1C'_1C_2C'_2JJ'}) \leq 10\eps .
\end{equation}
That is, $\Phi'_{RABC_1C'_1} \in \ball{10\eps}{\Psi_{RABC_1C'_1}}$. The number of bits communicated by Alice to Bob in $\cP$ is upper bounded by:
$$\log(n/b) \leq \dmaxeps{\Psi_{RBC}}{\Psi_{RB}\otimes \sigma_C}{\eps} - \dheps{\Psi_{BC}}{\Psi_{B}\otimes \sigma_C}{\eps^2} + 7\log\left(\frac{1}{\varepsilon}\right).$$ 

\vspace{0.1in}

\noindent {\bf Randomness required:} Let $\sigma_C$ be chosen to be $\frac{\id_C}{|C|}$, that is, the maximally mixed quantum state. By expanding the dimension of $\cH_C$ from $|C|$ to $2|C|$ if required, we can assume that $|C|$ is a prime (due to Bertrand's postulate \cite{Chebysev1852}). The shared randomness $\bsigma_{C_1C'_1C_2C'_2\ldots C_nC'_n}$ is to be chosen such that for all $i$, $\barq(c_i) = \frac{1}{|C|}$ and for all $i\neq j$, $\barq(c_i,c_j)= \frac{1}{|C|^2}$. We construct $\barq$ using Claim \ref{pairwisedist}. The number of bits of shared randomness required to generate $\barq$ is $$\left\lceil \frac{\log n}{\log |C|}\right\rceil\cdot \log|C|+ \log|C| \leq \max\left\{\log n , \log|C|\right\} + \log|C|.$$ 
From the choice of $n$, we have
$$\log n = \dmaxeps{\Psi_{RBC}}{\Psi_{RB}\otimes \frac{\id_C}{|C|}}{\eps} + \log\frac{8}{\eps^5} \leq \log|C| +  \log\frac{8}{\eps^5}.$$
This leads to the desired bound on the amount of randomness required.

\vspace{0.1in}

\noindent {\bf Randomness returned:} As concluded in Equation \ref{eq:finaldist}, $$\Pur\left(\Phi'_{RBAJJ'C_1C'_1C_2C'_2JJ'},\Psi_{RBAC_1C'_1}\otimes \left(\frac{1}{n}\sum_j \ketbra{j,j}_{JJ'}\right)\otimes \bsigma_{C_2C'_2}\right) \leq 10\eps.$$
Thus, $\log n +\log|C|$ bits of shared randomness are returned by the protocol with error at most $10\eps$  in purified distance. This completes the proof by using the value of $n$.
\end{proof}

\subsection*{Claims used in the theorem}
\noindent A few claims were used in the proof that we discuss below. Following claim is well known for classical quantum states, which we prove for completeness.
\begin{claim}[Fidelity between classical quantum states]
\label{fid:cq}
Let $\rho_{XA},\sigma_{XA}$ be two c-q states of the form $$\rho_{XA} = \sum_x p(x)\ketbra{x}_X\otimes \rho^x_A, \quad \sigma_{XA} = \sum_x q(x)\ketbra{x}_X\otimes \sigma^x_{A}.$$ Then 
$$\F(\rho_{XA},\sigma_{XA}) = \sum_x \sqrt{p(x)q(x)}\F(\rho^x_A,\sigma^x_A).$$
\end{claim}

\begin{proof}
We have that 
\begin{eqnarray*}
&&\F(\rho_{XA},\sigma_{XA})\\ &=& \Tr\bigg(\sqrt{\sum_x p(x)\ketbra{x}_X\otimes \rho^x_A}\sum_x q(x)\ketbra{x}_X\otimes \sigma^x_A  \sqrt{\sum_x p(x)\ketbra{x}_X\otimes \rho^x_A} \bigg)^{1/2}\\ &=&
\Tr\bigg((\sum_x \sqrt{p(x)}\ketbra{x}_X\otimes \sqrt{\rho^x_A})(\sum_x q(x)\ketbra{x}_X\otimes \sigma^x_A)(\sum_x \sqrt{p(x)}\ketbra{x}_X\otimes \sqrt{\rho^x_A}) \bigg)^{1/2} \\ &=&
\Tr\bigg(\sum_x p(x)q(x)\ketbra{x}_X\otimes \sqrt{\rho^x_A}\sigma^x_A\sqrt{\rho^x_A} \bigg)^{1/2} \\ &=&
\Tr\bigg(\sum_x \sqrt{p(x)q(x)}\ketbra{x}_X\otimes \sqrt{\sqrt{\rho^x_A}\sigma^x_A\sqrt{\rho^x_A}}\bigg) \\ &=&
\sum_x \sqrt{p(x)q(x)} \Tr\left(\sqrt{\sqrt{\rho^x_A}\sigma^x_A\sqrt{\rho^x_A}}\right) \\ &=& \sum_x \sqrt{p(x)q(x)}\F(\rho^x_A,\sigma^x_A).
\end{eqnarray*}
This proves the claim.
\end{proof}

We have used the following classical-quantum version of Uhlmann's theorem.

\begin{claim}[Classical-quantum Uhlmann's theorem]
\label{fact:cquhlmann}
Let $\rho_{XAB},\sigma_{XAC}$ be two c-q states of the form $$\rho_{XAB} = \sum_x p(x)\ketbra{x}_X\otimes \ketbra{\rho^x}_{AB}, \quad \sigma_{XAC} = \sum_x q(x)\ketbra{x}_X\otimes \ketbra{\sigma^x}_{AC}.$$ There exists a set of isometries $\{U^x:B\rightarrow C\}$ such that 
$$\F\left(\left(\sum_x\ketbra{x}_X\otimes  I_A\otimes U^x\right)\rho_{XAB}\left(\sum_x\ketbra{x}_X\otimes I_A\otimes U^{x\dagger}\right), \sigma_{XAC}\right) = \F(\rho_{XA},\sigma_{XA}).$$
\end{claim}

\begin{proof}
For every $x$, there exists an isometry $U_x:B\rightarrow C$, as guaranteed by Uhlmann's Theorem \ref{uhlmann}, such that $$\F\left((I_A\otimes U^x)\ketbra{\rho^x}_{AB}(I_A\otimes U^{x\dagger}), \ketbra{\sigma^x}_{AC}\right) = \F(\rho^x_A,\sigma^x_A).$$
The fact follows from the expression (Fact \ref{fid:cq}) $$\F(\rho_{XA},\sigma_{XA}) = \sum_x \sqrt{p(x)q(x)} \F(\rho^x_A,\sigma^x_A),$$ and the relation
\begin{eqnarray*}
&&\F\left(\left(\sum_x\ketbra{x}_X\otimes  I_A\otimes U^x\right)\rho_{XAB}\left(\sum_x\ketbra{x}_X\otimes I_A\otimes U^{x\dagger}\right), \sigma_{XAC}\right)\\ && = \sum_x \sqrt{p(x)q(x)} \F\left(( I_A\otimes U^x)\ketbra{\rho^x}_{AB}( I_A\otimes U^{x\dagger}),\ketbra{\sigma^x}_{AC}\right) .
\end{eqnarray*}
\end{proof}

Gentle measurement lemma is used to prove the following claim applied in conjunction with hypothesis testing.

\begin{claim}[Pretty good POVM]
\label{gentlepovm}
Consider a quantum state $\rho_A = \sum_i \lambda_i \rho^i_A$ and a map $\cA (X) = \sum_i P_i X P_i \otimes \ketbra{i}_O$, such that $0 < P_i < I, \sum_i P_i^2 = I$ ($O$ is considered the output register for the measurement $\cA$). Define the state $\rho'_{AO}\defeq \sum_i p_i\rho^i_A\otimes \ketbra{i}_O$ and let $p_{i|j} \defeq \Tr(P^2_i \rho^j_A)$ be the probability of obtaining outcome $i$ on quantum state $\rho^j_A$. Then it holds that
$$\F(\rho'_{AO},\cA(\rho_A))\geq (\sum_i \lambda_ip_{i|i})^{3/2}.$$
\end{claim}

\begin{proof}
We abbreviate $\sigma_{AO} \defeq \cA(\rho_A)$. This implies that
$$\sigma_{AO} = \sum_{i,j} \lambda_i P_j\rho^i_A P_j \otimes \ketbra{j}_O.$$
Define $$\sigma^{good}_{AO} \defeq \frac{1}{\sum_i \lambda_ip_{i|i}}\sum_{i} \lambda_i P_i\rho^i_A P_i \otimes \ketbra{i}_O\quad\mbox{and}\quad \sigma^{bad}_{AO} \defeq \frac{1}{1-\sum_i \lambda_ip_{i|i}}\sum_{i\neq j} \lambda_i P_j\rho^i_A P_j \otimes \ketbra{j}_O.$$ Then we can decompose $\sigma_{AO}$ as 
$$\sigma_{AO} = (\sum_i \lambda_ip_{i|i})\sigma^{good}_{AO} + (1-\sum_i\lambda_ip_{i|i}) \sigma^{bad}_{AO}. $$  From concavity of fidelity, this gives us 
\begin{align*}
\F(\sigma_{AO}, \rho'_{AO}) &\geq (\sum_i \lambda_ip_{i|i})\F(\sigma^{good}_{AO},\rho'_{AO}) \\
&= (\sum_i \lambda_ip_{i|i})\cdot \left(\frac{1}{\sqrt{\sum_i \lambda_ip_{i|i}}}\sum_i \lambda_i\sqrt{p_{i|i}}\cdot \F(\frac{P_i\rho^i_A P_i}{p_{i|i}},\rho^i_A)\right).
\end{align*} 
Now employing Gentle measurement lemma (Fact \ref{gentlelemma}), we conclude that
$$\F(\sigma_{AO}, \rho'_{AO}) \geq  \sqrt{\sum_i \lambda_ip_{i|i}}(\sum_i \lambda_i\sqrt{p_{i|i}}\cdot \sqrt{p_{i|i}}) = (\sum_i \lambda_ip_{i|i})^{3/2}.$$ 
\end{proof}

\subsection{A converse bound}
\label{subsec:convbound}

The work \cite{LeditzkyWD16} provided some converse bounds for the task in Definition \ref{def:qmc} in terms of quantum R\'{e}nyi entropies. Here, we provide converse bounds in terms of smooth max-relative entropy. The converse below closely follows the proof of converse for quantum state redistribution given in \cite{Berta14}.

\begin{corollary}
\label{cor:converse}
Fix $\eps, \delta\in (0,1)$, and consider the state $\ketbra{\Psi}_{RAC\bar{C}B}$ of the form $\sum_c \sqrt{p(c)} \ket{c}_C \ket{c}_{\bar{C}} \ket{\psi^c}_{RAB}$. For any $(m, r_1, r_2, \eps)$- quantum measurement compression protocol for $\ketbra{\Psi}_{RAC\bar{C}B}$, it holds that
$$m \geq \inf_{\sigma_{BC}}\dmaxeps{\Psi_{RBC}}{\Psi_R\otimes \sigma_{BC}}{\eps+\delta} - \dmaxeps{\Psi_{RB}}{\Psi_R\otimes \Psi_B}{\delta}.$$
\end{corollary}
\begin{proof}
Let $\Psi'_{RB}$ be the quantum state achieving the optimum in $\dmaxeps{\Psi_{RB}}{\Psi_R\otimes \Psi_B}{\delta}$. Let $\ketbra{\Psi'}_{RAB}$ be it's purification such that $\Pur(\Psi'_{RAB}, \Psi_{RAB}) \leq \delta$ (as guaranteed by Uhlmann's theorem, Fact \ref{uhlmann}). Suppose Alice measures $\Psi'_{RAB}$ obtaining outcomes in register $C\bar{C}$ and runs the protocol on the resulting quantum state. Let $\Omega_{RBQE_B}$ be the quantum state with Bob after Alice's message. Since $QE_B$ are classical registers, we have
$$\Omega_{RBQE_B} \preceq |Q|\Psi'_{RB}\otimes \frac{\id_Q}{|Q|}\otimes \theta_{E_B},$$ as $\Omega_{RBE_B} = \Psi'_{RB}\otimes\theta_{E_B}$. Thus,  
$$\Omega_{RBQE_B} \preceq |Q|2^{\dmaxeps{\Psi_{RB}}{\Psi_R\otimes \Psi_B}{\delta}}\Psi_{R}\otimes \Psi_B\otimes \frac{\id_Q}{|Q|}\otimes \theta_{E_B}.$$
Applying Bob's operation on $\Omega_{RBQE_B}$ and tracing out register $T_B$, we obtain a quantum state $\Psi''_{RBC'}$ such that $\Pur(\Psi''_{RBC'}, \Psi_{RBC'}) \leq \eps+\delta$. Let $\sigma_{BC}$ be the quantum state obtained after applying Bob's operation on $\Psi_B\otimes \frac{\id_Q}{|Q|}\otimes \theta_{E_B}$ and tracing out register $T_B$. Then,
$$\Psi''_{RBC'} \preceq |Q|2^{\dmaxeps{\Psi_{RB}}{\Psi_R\otimes \Psi_B}{\delta}}\Psi_{R}\otimes \sigma_{BC}.$$
Thus, we conclude that
$$\inf_{\sigma_{BC}}\dmaxeps{\Psi_{RBC}}{\Psi_R\otimes \sigma_{BC}}{\eps+\delta} \leq \log|Q| + \dmaxeps{\Psi_{RB}}{\Psi_R\otimes \Psi_B}{\delta}.$$ This completes the proof by setting $m=\log|Q|$.
\end{proof}

We point out that combining the converse in \cite{LeditzkyWD16} and the relations between quantum R\'{e}nyi entropies and smooth conditional min-entropy \cite[Corollary 3.6]{MillerShi16}, we conclude that the following is a converse bound on the classical communication cost:
$$\hmineps{R}{B}{\eps}_{\Psi} - \hmaxeps{R}{CB}{\eps}_{\Psi} - O\left(\log\frac{1}{\eps}\right).$$
This expression may be incomparable to that obtained in Corollary \ref{cor:converse}.

\subsection{Asymptotic and i.i.d. analysis}

Now, we discuss the asymptotic and i.i.d. behavior of our bounds, showing the randomness required and communication cost of the protocol in Theorem \ref{newcompression}. We show the following theorem, where we use the shorthand $R^n$ (similarly for other registers) to represent $n$ copies of the register $R$. This result is obtained by running the protocol obtained in Theorem \ref{newcompression} several times and recycling the shared randomness each time.
\begin{theorem}
\label{asympmeascomp}
Let $\ket{\Psi}_{RAC\bar{C}B} := \sum_c \sqrt{p(c)} \ket{c}_C \ket{c}_{\bar{C}} \ket{\psi^c}_{RAB}$ be a quantum state. For every $\eps \in (0,1/10)$ and integer $n\geq 1$, there exists a $(Q(n,\eps), S(n, \eps), O(\sqrt{n}), 10\eps)$- quantum measurement compression protocol for the quantum state $\ket{\Psi}_{RABC\bar{C}}^{\otimes n}$ such that 
$$\lim_{n\rightarrow \infty}\frac{1}{n} Q(n,\eps) = \condmutinf{R}{C}{B}, \lim_{n\rightarrow \infty}\frac{1}{n} S(n,\eps) = \mathrm{H}(C|BR).$$
\end{theorem}

\begin{proof}
Let $m:= \sqrt{n}$ and $\eps':= \frac{m\eps}{n}$. We divide the state $\ket{\Psi}_{RAC\bar{C}B}^{\otimes n}$ into $\frac{n}{m}$ blocks of states $\ket{\Psi}_{RAC\bar{C}B}^{\otimes m}$. Alice and Bob pre-share $$2m\log|C| - \frac{n}{m}\dmaxeps{\Psi_{RBC}^{\otimes m}}{\Psi_{RB}^{\otimes m}\otimes \id_C^{\otimes m}}{\eps'} - 5\log \eps'$$ bits of shared randomness. Consider,
\begin{eqnarray*}
\dmaxeps{\Psi_{RBC}^{\otimes m}}{\Psi_{RB}^{\otimes m}\otimes \frac{\id_C^{\otimes m}}{|C|^m}}{\eps'} &=& m\log|C| + \dmaxeps{\Psi_{RBC}^{\otimes m}}{\Psi_{RB}^{\otimes m}\otimes \id_C^{\otimes m}}{\eps'} \\ &\overset{a}=& m\log|C| + m\relent{\Psi_{RBC}}{\Psi_{RB}\otimes \id_C} \\ &+& \sqrt{m \varrelent{\Psi_{RBC}}{\Psi_{RB}\otimes \id_C}}\Phi^{-1}(\eps') + O(\log m) \\ &=&  m\log|C| - m \mathrm{H}(C|RB)_{\Psi} + \sqrt{m \varrelent{\Psi_{RBC}}{\Psi_{RB}\otimes \id_C}}\Phi^{-1}(\eps') + O(\log m),
\end{eqnarray*}
where (a) uses Fact \ref{dmaxequi}. Combining this with the identity $$\dmaxeps{\Psi_{RBC}^{\otimes m}}{\Psi_{RB}^{\otimes m}\otimes \frac{\id_C^{\otimes m}}{|C|^m}}{\eps'} = \dmaxeps{\Psi_{RBC}^{\otimes m}}{\Psi_{RB}^{\otimes m}\otimes \id_C^{\otimes m}}{\eps'} + m\log|C|,$$
we find that the number of bits of randomness initially present with Alice and Bob is 
\begin{eqnarray}
\label{sharedrandom}
S(n,\eps) &=& 2m\log|C| - \frac{n}{m}\dmaxeps{\Psi_{RBC}^{\otimes m}}{\Psi_{RB}^{\otimes m}\otimes \id_C^{\otimes m}}{\eps'} - 5\log \eps' \nonumber\\ &=& 2m\log|C| + n\mathrm{H}(C|RB)_{\Psi} + \frac{n}{\sqrt{m}}\sqrt{\varrelent{\Psi_{RBC}}{\Psi_{RB}\otimes \id_C}}\Phi^{-1}(\eps') + O(\log (mn/\eps)) \nonumber \\ &\overset{a}\leq&  
2m\log|C| + n\mathrm{H}(C|RB)_{\Psi} + \frac{n\sqrt{\log \frac{n}{m\eps}}}{\sqrt{m}}\sqrt{\varrelent{\Psi_{RBC}}{\Psi_{RB}\otimes \id_C}} + O(\log (mn/\eps)),
\end{eqnarray}
where (a) uses Fact \ref{gaussianupper}. Similarly, we have
\begin{equation}
\label{sharedrandomlower}
S(n,\eps) \geq 2m\log|C| + n\mathrm{H}(C|RB)_{\Psi} - \frac{n\sqrt{\log \frac{n}{m\eps}}}{\sqrt{m}}\sqrt{\varrelent{\Psi_{RBC}}{\Psi_{RB}\otimes \id_C}} + O(\log (mn/\eps))
\end{equation}
Alice and Bob run protocol $\cP$ given in Theorem \ref{newcompression} on the quantum state $\Psi_{RAC\bar{C}B}^{\otimes m}$ with $2m\log|C| - 5\log\eps'$ bits of shared randomness. The protocol $\cP$ returns $$\dmaxeps{\Psi_{RBC}^{\otimes m}}{\Psi_{RB}^{\otimes m}\otimes \frac{\id_C^{\otimes m}}{|C|^m}}{\eps'} + m\log|C| = 2m\log|C| + \dmaxeps{\Psi_{RBC}^{\otimes m}}{\Psi_{RB}^{\otimes m}\otimes \id_C^{\otimes m}}{\eps'}$$ bits of shared randomness with error $10\eps'$. Thus, the protocol consumes $- \dmaxeps{\Psi_{RBC}^{\otimes m}}{\Psi_{RB}^{\otimes m}\otimes \id_C^{\otimes m}}{\eps'}$ bits of shared randomness.  After this, Alice and Bob run the protocol for second block. For this, they add $- \dmaxeps{\Psi_{RBC}^{\otimes m}}{\Psi_{RB}^{\otimes m}\otimes \id_C^{\otimes m}}{\eps'}$ bits of shared randomness (to compensate for the amount consumed in previous protocol; note that this quantity is positive, by Fact \ref{dmaxwithid}). This process continues for all the $\frac{n}{m}$ blocks.

From Fact \ref{slowchange}, we have that the overall error in terms of purified distance is at most $10\frac{n}{m}\eps' \leq 10\eps$. The number of bits communicated in the protocol is 
\begin{eqnarray*}
Q(n,\eps)&=&\frac{n}{m}\bigg(\dmaxeps{\Psi_{RBC}^{\otimes m}}{\Psi_{RB}^{\otimes m} \otimes \frac{\id_C^{\otimes m}}{|C|^m}}{\eps'} - \dheps{\Psi^{\otimes m}_{BC}}{\Psi^{\otimes m}_{B}\otimes \frac{\id_C^{\otimes m}}{|C|^m}}{\eps'^2} + 6\log\left(\frac{1}{\eps'}\right)\bigg) \\ &=& \frac{n}{m}\bigg(\dmaxeps{\Psi_{RBC}^{\otimes m}}{\Psi_{RB}^{\otimes m} \otimes \id_C^{\otimes m}}{\eps'} - \dheps{\Psi^{\otimes m}_{BC}}{\Psi^{\otimes m}_{B}\otimes \id_C^{\otimes m}}{\eps'^2} + 6\log\left(\frac{1}{\eps'}\right)\bigg)\\
&\overset{a}\leq& \frac{n}{m}\bigg(m \relent{\Psi_{RBC}}{\Psi_{RB}\otimes \id_C} + \sqrt{m\varrelent{\Psi_{RBC}}{\Psi_{RB}\otimes \id_C}}|\Phi^{-1}(\eps')| + O(\log m) \\ &-& m\relent{\Psi_{BC}}{\Psi_B\otimes \id_C} + \sqrt{m\varrelent{\Psi_{BC}}{\Psi_{B}\otimes \id_C}}|\Phi^{-1}(\eps')| + 6\log\left(\frac{1}{\eps'}\right)\bigg) \\ &=& n\cdot \condmutinf{R}{C}{B}_{\Psi} + \frac{n}{m}O(\log (mn/\eps)) \\ &+& \frac{n}{\sqrt{m}}(\sqrt{\varrelent{\Psi_{BC}}{\Psi_{B}\otimes \id_C}} + \sqrt{\varrelent{\Psi_{RBC}}{\Psi_{RB}\otimes \id_C}})|\Phi^{-1}(\eps')|\\ &\overset{b}\leq& n\cdot \condmutinf{R}{C}{B}_{\Psi} + \frac{n}{m}O(\log (mn/\eps)) \\ &+& \frac{n\sqrt{\log \frac{n}{m\eps}}}{\sqrt{m}}(\sqrt{\varrelent{\Psi_{BC}}{\Psi_{B}\otimes \id_C}} + \sqrt{\varrelent{\Psi_{RBC}}{\Psi_{RB}\otimes \id_C}}),
\end{eqnarray*}
where (a) uses Fact \ref{dmaxequi} and (b) uses Fact \ref{gaussianupper}. Similarly, we have
\begin{eqnarray*}
Q(n,\eps)&\geq& n\cdot \condmutinf{R}{C}{B}_{\Psi} + \frac{n}{m}O(\log (mn/\eps)) - \frac{n\sqrt{\log \frac{n}{m\eps}}}{\sqrt{m}}(\sqrt{\varrelent{\Psi_{BC}}{\Psi_{B}\otimes \id_C}} + \sqrt{\varrelent{\Psi_{RBC}}{\Psi_{RB}\otimes \id_C}})
\end{eqnarray*}
Combining with Equations \ref{sharedrandom}, \ref{sharedrandomlower} and setting $m=\sqrt{n}$, we find that 
$$\lim_{n\rightarrow \infty} \frac{1}{n}S(n,\eps) = \mathrm{H}(C|RB)_{\Psi}, \quad \lim_{n\rightarrow \infty} \frac{1}{n} Q(n,\eps) = \condmutinf{R}{C}{B}_{\Psi}.$$  
This completes the proof.
\end{proof}

\section{Randomness extraction and quantum measurement compression without feedback}
\label{sec:randext}

We formally define the task of a quantum proof (strong) randomness extraction, adapted from \cite[Definition 3.2]{DePVR12}. 
\begin{definition}[A quantum proof strong randomness extractor]
\label{def:randext}
Fix an $\eps \in (0,1)$ and a register $C$ with a basis $\{\ket{c}\}_{c=1}^{|C|}$. Let $\theta_U = \frac{\id_U}{|U|}$ be a uniform distribution on register $U$ (the seed). A $(k, \log|U|, \log|V|, \eps)$ - quantum proof randomness extraction protocol consists of a register $V$, a unitary $W: \cH_C\otimes \cH_A \otimes \cH_U \rightarrow \cH_{C'}\otimes\cH_V\otimes \cH_{\bar{U}}$ for some ancilla register $A$, register $C'$ and $\bar{U} \equiv U$. It is required that for all classical-quantum states $\Psi_{GC}$, where $C$ is the classical register in the basis $\{\ket{c}\}_{c=1}^{|C|}$, satisfying $\dmax{\Psi_{GC}}{\Psi_G \otimes \id_C}\leq -k$,
$$\relent{\Tr_{C'}W(\Psi_{GC}\otimes \ketbra{0}_A\otimes \theta_U)W^{\dagger}}{\Psi_G\otimes \frac{\id_V}{|V|}\otimes\theta_{\bar{U}}} \leq \eps.$$ 
\end{definition}

\begin{remark}
\label{rem:condindef}
{\bf Conditions in the  above definition:} The error in above definition is measured in terms of relative entropy, as opposed to trace distance in \cite{DePVR12}, making our criteria stronger. The conditional min-entropy criteria $\hmin{C}{G}_{\Psi} \geq k$ in \cite{DePVR12} is weakened to $-\dmax{\Psi_{GC}}{\Psi_G\otimes\id_C}\geq k$. But this weakening does not lead to much difference if one measures the error in trace distance (as done in \cite{DePVR12}) and allows a further error of $\delta \in (0,1)$. For this, use Fact \ref{imaxvariants} to observe that
$$-\min_{\Psi'_{GC}\in \ball{\delta}{\Psi_{GC}}}\dmax{\Psi'_{GC}}{\Psi'_G\otimes \id_C} \geq -\min_{\sigma_G}\dmaxeps{\Psi_{GC}}{\sigma_G\otimes \id_C}{\frac{\delta}{2}} - 3\log\frac{4}{\delta} =  \hmineps{C}{G}{\frac{\delta}{2}}_{\Psi} - 3\log\frac{4}{\delta}.$$
Thus, for any quantum state $\Psi_{GC}$ satisfying $\hmineps{C}{G}{\frac{\delta}{2}}_{\Psi} \geq k + 3\log\frac{4}{\delta}$, there exists a quantum state $\Psi'_{GC}\in \ball{\delta}{\Psi_{GC}}$ (which can be assume to be classical-quantum, see Corollary \ref{corconvex}) such that $-\dmax{\Psi'_{GC}}{\Psi'_G\otimes \id_C}\geq k$. A $(k, \log|U|, \log|V|, \eps)$ - quantum proof randomness extractor (with associated unitary $W$) satisfies
$$\relent{\Tr_{C'}W(\Psi'_{GC}\otimes \ketbra{0}_A\otimes \theta_U)W^{\dagger}}{\Psi'_G\otimes \frac{\id_V}{|V|}\otimes\theta_{\bar{U}}} \leq \eps.$$ 
Fact \ref{pinsker} implies that
$$\Pur\left(\Tr_{C'}W(\Psi'_{GC}\otimes \ketbra{0}_A\otimes \theta_U)W^{\dagger},\Psi'_G\otimes \frac{\id_V}{|V|}\otimes\theta_{\bar{U}}\right) \leq \sqrt{\eps}.$$ 
Using triangle inequality for purified distance (Fact \ref{fact:trianglepurified}), we conclude that 
$$\Pur\left(\Tr_{C'}W(\Psi_{GC}\otimes \ketbra{0}_A\otimes \theta_U)W^{\dagger},\Psi_G\otimes \frac{\id_V}{|V|}\otimes\theta_{\bar{U}}\right) \leq \sqrt{\eps} + 2\delta.$$ 
From Fact \ref{fuchsG}, it holds that
$$\Delta\left(\Tr_{C'}W(\Psi_{GC}\otimes \ketbra{0}_A\otimes \theta_U)W^{\dagger},\Psi_G\otimes \frac{\id_V}{|V|}\otimes\theta_{\bar{U}}\right) \leq \sqrt{\eps} + 2\delta.$$ 
Thus, $\log|V|$ bits are extracted with error $\sqrt{\eps}+2\delta$ in trace distance, whenever $\hmineps{C}{G}{\frac{\delta}{2}}_{\Psi} \geq k + 3\log\frac{4}{\delta}$.
\end{remark}

\begin{remark}
\label{rem:srandext}
{\bf  Shared randomness extraction:} A protocol captured by Definition \ref{def:randext} can also be used to obtain shared randomness in two-party setting. More precisely, if the joint quantum state $\Psi_{GCC'} = \sum_c p(c) \ketbra{c,c}_{CC'}\otimes \Psi^c_G$ is shared in a manner that Alice holds $C$, Bob holds $C'$ and $G$ is shared between Alice, Bob and a third party, then Alice and Bob can jointly run a quantum proof strong randomness extractor protocol, using the shared randomness $\theta_{UU'}= \frac{1}{|U|}\sum_u\ketbra{u,u}_{U,U'}$ and gain a shared randomness $\tau_{VV'} = \frac{1}{|V|}\sum_v \ketbra{v,v}_{V,V'}$. It holds that $\tau_{VV'}\otimes \theta_{UU'}$ is almost independent of $\Psi_G$, up to error $\eps$ in purified distance.
\end{remark}
We have the following result.  
\begin{theorem}
\label{randext}
For $\eps\in (0,1)$, a register $C$ with a basis $\{\ket{c}\}_{c=1}^{|C|}$ and a real number $k>0$, there exists a $$\left(k, 2\log|C| - k + 2\log\frac{1}{\eps}, k - \log\frac{1}{\eps} - 1, \eps\right)$$ - quantum proof randomness extraction protocol.
\end{theorem}

\begin{remark}
\label{rem:efficient}
{\bf Efficiency of the protocol:} While we only provide an achievability proof in the information theoretic sense, it can be observed that the protocol is also efficient. This follows from the fact that the construction of pairwise independent random variables (as stated in Fact \ref{pairwise}, derived from \cite{Lovettnotes}) can be done efficiently.
\end{remark}

\begin{proof}[Proof of Theorem \ref{randext}]
Let $n\defeq \lceil \frac{|C|\cdot 2^{-k}}{\eps}\rceil$. Let $\barq$ be the pairwise independent distribution over $\cC\times \cC \times \ldots \cC$ ($n$ times) as constructed in Claim \ref{pairwisedist}. Let $C_1, \ldots C_n \equiv C$ be $n$ copies of the register $C$ and $\sigma_{C_1, \ldots C_n}$ be the quantum state obtained from the distribution $\barq$ in the basis $\{\ket{c_1, \ldots c_n}\}$.  For any quantum state $\Psi_{GC}$ satisfying 
$$\dmax{\Psi_{GC}}{\Psi_G \otimes \frac{\id_C}{|C|}} = \log|C|+\dmax{\Psi_{GC}}{\Psi_G \otimes \id_C} \leq \log|C|-k,$$
invoke Lemma \ref{convsplitpairwise}, with $P \leftarrow G$ and $Q\leftarrow C$. From Lemma \ref{corconvex}, it holds that
\begin{equation}
\label{eq:randextconv}
\relent{\frac{1}{n}\sum_{j=1}^n\sum_c p(c)\Psi^c_G \otimes \ketbra{c}_{C_j}\otimes \sigma^{c}_{C_1, \ldots C_{j-1}, C_{j+1}, \ldots C_n}}{\Psi_G \otimes \sigma_{C_1, \ldots C_n}} \leq \eps.
\end{equation}
Set $b\defeq \lceil\frac{\log n}{\log |\cC|}\rceil$, let $U_1\defeq C\times C\times \ldots C$ ($b$ times) and $J$ be a register of dimension $n$. Let $U\defeq U_1J$ and $\theta_U= \frac{\id_{U_1}}{|U_1|}\times \frac{\id_J}{|J|}$. 

\vspace{0.1in}

\noindent {\bf Protocol:} The protocol is as follows, which is constructed only using the value $k$. 
\begin{enumerate}
\item Rename the reigster $C$ with $C_1$. Introduce registers $C_2, \ldots C_n$ in the state $\otimes_{i=2}^n\ketbra{1}_{C_i}$ and the register $U\equiv U_1J$ in the state $\theta_U$.
\item Conditioned on value $j$ in register $J$, swap $C_j$ with $C_1$. The global quantum state at this stage is
$$\frac{1}{n}\sum_{j=1}^n\ketbra{j}_J \otimes \Psi_{GC_j} \otimes \ketbra{1}_{C_1}\otimes \ldots \ketbra{1}_{C_{j-1}}\otimes \ketbra{1}_{C_{j+1}}\otimes \ldots \ketbra{1}_{C_n}\otimes \frac{\id_{U_1}}{|U_1|}.$$
\item Conditioned on the value $j$ in register $J$ and $c_j$ in register $C_j$, apply a unitary $W_1$ which maps the state
$$\ketbra{1}_{C_1}\otimes \ldots \ketbra{1}_{C_{j-1}}\otimes \ketbra{1}_{C_{j+1}}\otimes \ldots \ketbra{1}_{C_n}\otimes \frac{\id_{U_1}}{|U_1|}$$ to the state
$$\sigma^{c_j}_{C_1, \ldots C_{j-1}C_{j+1}\ldots C_n} \otimes \ketbra{1}_{U_1}.$$
This is possible since $\barq(c_1, \ldots c_{j-1}, c_{j+1}, \ldots c_n | c_j)$ is uniform in a support of size $|U_1| = |\cC|^{b}$, as guaranteed by Claim \ref{pairwisedist}. The global quantum state at this stage is 
$$\frac{1}{n}\sum_{j=1}^n\ketbra{j}_J\otimes\sum_c p(c)\Psi^c_G \otimes \ketbra{c}_{C_j}\otimes \sigma^{c}_{C_1, \ldots C_{j-1}, C_{j+1}, \ldots C_n}\otimes \ketbra{1}_{U_1}.$$  
\item Apply a unitary $W_2$ which maps the quantum state $$\sigma_{C_1, \ldots C_n}$$ to the quantum state $$ \frac{\id_{C_1}}{|\cC|} \otimes \ldots \frac{\id_{C_{b+1}}}{|\cC|}\otimes \ketbra{1}_{C_{b+2}}\otimes \ldots \ketbra{1}_{C_n}.$$ This is possible since $\barq$ is uniform in a support of size $|\cC|^{b+1}$, as guaranteed by Claim \ref{pairwisedist}. From Equation \ref{eq:randextconv}, the overall quantum state on registers $GC_1, \ldots C_n, U_1$ is close to 
$$\Psi_G \otimes \frac{\id_{C_1}}{|\cC|} \otimes \ldots \frac{\id_{C_{b+1}}}{|\cC|}\otimes \ketbra{1}_{C_{b+2}}\otimes \ldots \ketbra{1}_{C_n} \otimes \ketbra{1}_{U_1}.$$
\item Set $C'\defeq JC_{b+1}\ldots C_n U_1$ and choose $\bar{U}, V$ such that $\bar{U}V\defeq C_1\ldots C_{b+1}$ and $|\bar{U}|=|U| = |U_1||J|$. This can be achieved by dividing $C_1, \ldots C_{b+1}$ into smaller registers.
\end{enumerate}

\vspace{0.1in}

\noindent {\bf Analysis:} We obtain $$|V| = \frac{|\cC|^{b+1}}{|U_1||J|} = \frac{|\cC|^{b+1}}{|\cC|^b|J|} = \frac{|\cC|}{n}.$$ Thus, 
\begin{eqnarray*}
\log|V| &=& \log\frac{|\cC|}{n} \geq \log (2^k\eps) -1 = k - \log\frac{1}{\eps} -1.
\end{eqnarray*}
Further,
$$\log|U| = \log|U_1| + \log|J| = b\log|C| + \log n \leq 2\log|C| - k + 2\log\frac{1}{\eps},$$ where the inequality holds since $\log n = \log|C| -k +\log\frac{1}{\eps}$ which implies that $\lceil \frac{\log n}{\log|C|}\rceil\cdot \log|C| \leq \log|C| + \log\frac{1}{\eps}$. This completes the proof.
\end{proof}

\noindent{\bf Comparison with previous work:} It was shown in \cite{RadhakrishnanT00} that a randomness extractor acting on a source with min entropy $k$ can extract uniform distribution (up to error $\eps$ in trace distance) on at most $k - 2\log\frac{1}{\eps} + O(1)$ number of bits. This bound is achieved, up to additive constants, in \cite[Corollary 5.5.2]{Renner05}. The construction in \cite[Corollary 5.4]{DePVR12} extracts $k - 4\log\frac{1}{\eps}$ bits, but with exponential improvement in the seed size in comparison to Theorem \ref{randext} or \cite[Corollary 5.5.2]{Renner05}.  In our construction in Theorem \ref{randext}, error of $\eps$ in relative entropy allows us to extract $k - \log\frac{1}{\eps} -1$ number of uniform bits. By Fact \ref{pinsker}, error of $\eps$ in relative entropy implies an error of $2\sqrt{\eps}$ in trace distance, showing the optimality of our construction in terms of the number of bits extracted.

\subsection{Quantum measurement compression without feedback}

A consequence of our result on randomness extraction is that quantum measurement compression can be performed in the case where Alice does not need to possess the outcome of the measurement. We formally introduce the task.
\begin{definition}[Quantum measurement compression without feedback]
\label{def:qmcfed}
Fix an $\eps\in (0,1)$, and consider the state $\ketbra{\Psi}_{RA_0B}$. Let $\cN: \cL(A_0) \rightarrow \cL(AC)$ be a quantum measurement acting as $$\cN(\rho_{A_0}) = \sum_c \ketbra{c}_C\otimes N_c\rho_{A_0} N^{\dagger}_c$$ and $\Psi_{RAC} \defeq \cN(\ketbra{\Psi}_{RA_0B})$. An $(m, r_1, r_2, \eps)$- quantum measurement compression protocol without feedback consists of 
\begin{itemize}
\item a classical-classical state (or preshared randomness) $\theta_{E_AE_B}$ between Alice ($E_A$) and Bob ($E_B$), 
\item an encoding map by Alice $\cE: \cL(A_0E_A) \rightarrow \cL(QT_A)$, and
\item a decoding map by Bob $\cD: \cL(QBE_B)\rightarrow \cL(BC'T_B)$, where $C'\equiv C$
\end{itemize}
such that 
$$\Pur\left(\cD\cE(\Psi_{RABC}\otimes \theta_{E_AE_B}),\Psi_{RBC'}\otimes \omega_{T_AT_B}\right)\leq \eps,$$ for some classical-classical state $\omega_{T_AT_B}$.
The number of bits communicated is $m=\log|Q|$, number of bits of initial shared randomness is $r_1= \max(\log|E_A|, \log|E_B|)$ and final number of bits of shared randomness is $r_2 = \min(\log|T_A|, \log|T_B|)$. 
\end{definition}

We have the following one-shot result. 
\begin{theorem}
\label{theo:meascompfed}
Fix $\eps, \delta\in (0,1)$, a quantum state $\ketbra{\Psi}_{RA_0B}$ and a quantum measurement $\cN:\cL(A_0) \rightarrow \cL(AC)$. Fix a register $W$. Let $\cM: \cL(A_0)\rightarrow \cL(AW)$ be a quantum measurement acting as $$\cM(\rho_{A_0}) = \sum_w \ketbra{w}_W\otimes M_w\rho_{A_0}M^{\dagger}_w$$ and $p(c | w)$ be a probability distribution conditioned on each $w$, such that $$\cN(\rho_{A_0}) = \sum_{c,w} p(c|w) \ketbra{c}_C\otimes M_w\rho_{A_0}M^{\dagger}_w.$$ Define the quantum state $$\Psi_{RBACW}\defeq \sum_{c,w} p(c|w) \ketbra{c}_C\otimes \ketbra{w}_W\otimes M_w\Psi_{RBA_0}M^{\dagger}_w.$$ There exists a $(m,r_1, r_2, 10\eps+3\delta)$ - quantum measurement compression protocol without feedback such that
$$m\leq \dmaxeps{\Psi_{RBW}}{\Psi_{RB}\otimes \id_W}{\eps} - \dheps{\Psi_{BW}}{\Psi_{B}\otimes \id_W}{\eps^2} + 7\log\left(\frac{1}{\varepsilon}\right),$$
$$r_1 \leq 4\log|W| + \log\frac{64}{\eps^5\delta^5}$$ 
and
$$r_2 \geq 4\log|W| + \dmaxeps{\Psi_{RBW}}{\Psi_{RB}\otimes \id_W}{\eps}-  \dmaxeps{\Psi_{RBCW}}{\Psi_{RBC}\otimes \id_W}{\frac{\delta}{2}} -  \log\frac{8\eps^5}{\delta^5}.$$
\end{theorem}
\begin{proof}
Let $U_{\cM}: \cH_{A_0} \rightarrow \cH_A \otimes \cH_W \otimes \cH_{\bar{W}}$ be the isometry acting as 
$$U_{\cM}\ket{\rho}_{A_0} = \sum_w \ket{w,w}_{W\bar{W}}\otimes M_w\ket{\rho}_{A_0}.$$ Define $\ket{\Psi}_{RBAW\bar{W}}\defeq U_{\cM}\ket{\Psi}_{RBA_0}$. Observe that $\Psi_{RBAW} = \sum_w p(w)\ketbra{w}_W\otimes \Psi^w_{RBA}$, for some probability distribution $p(w)$. Alice and Bob implement the $(m, r'_1, r'_2, 10\eps)$- quantum measurement compression protocol as given in Theorem \ref{newcompression} with $\sigma_W = \frac{\id_W}{|W|}$, 
$$m\leq \dmaxeps{\Psi_{RBW}}{\Psi_{RB}\otimes \id_W}{\eps} - \dheps{\Psi_{BW}}{\Psi_{B}\otimes \id_W}{\eps^2} + 7\log\left(\frac{1}{\varepsilon}\right),$$
$$ r'_1\leq 2\log|W| + \log\frac{8}{\eps^5}$$ and
 $$r'_2 \geq 2\log|W| + \dmaxeps{\Psi_{RBW}}{\Psi_{RB}\otimes \id_W}{\eps} +  \log\frac{8}{\eps^5} .$$
The protocol outputs the quantum state $\Psi_{RBAWW'}$ up to error $10\eps$ in purified distance, where Alice holds the register $W$ and Bob holds the register $W'$. Since $\Psi_{RBAWW'} = \sum_w p(w)\ketbra{w, w}_{W, W'}\otimes \Psi^w_{RBA}$, Bob can produce the register $C'$ according to the conditional distribution $p(c|w)$. Then the resulting joint distribution is
$$\Psi_{RBAWW'C'} = \sum_{w,c} p(c|w)p(w)\ketbra{c}_{C'}\otimes\ketbra{w, w}_{W, W'}\otimes \Psi^w_{RBA}.$$
Let $k:= -\min_{\Psi'_{RBCW}\in \ball{\delta}{\Psi_{RBCW}}}\dmax{\Psi'_{RBCW}}{\Psi'_{RBC}\otimes \id_W}$. Following Remark \ref{rem:srandext}, Alice and Bob run the $$(k, 2\log|W| - k - 4\log\frac{1}{\delta}, k-2\log\frac{1}{\delta}, \delta^2)$$ -quantum proof randomness extraction protocol as obtained in Theorem \ref{randext}, with number of initial shared randomness equal to 
$$2\log|W| - k - 4\log\frac{1}{\delta},$$ such that the number of bits of shared randomness gained by the protocol is at least
$$k - \log\frac{1}{\delta^2} \geq -\dmaxeps{\Psi_{RBCW}}{\Psi_{RBC}\otimes \id_W}{\frac{\delta}{2}} - \log\frac{64}{\delta^5}.$$
This implies that 
total initial randomness required for the protocol is 
$$r_1 \leq 4\log|W| + \log\frac{64}{\eps^5\delta^5}$$ 
and the total final randomness obtained is
$$r_2 \geq 4\log|W| + \dmaxeps{\Psi_{RBW}}{\Psi_{RB}\otimes \id_W}{\eps}-  \dmaxeps{\Psi_{RBCW}}{\Psi_{RBC}\otimes \id_W}{\frac{\delta}{2}} -  \log\frac{8\eps^5}{\delta^5}.$$
The overall error in purified distance is $10\eps+3\delta$, by applying triangle inequality for purified distance (Fact \ref{fact:trianglepurified}) to the error guarantees in Theorems \ref{newcompression} and \ref{randext} (we also apply Fact \ref{pinsker} to Theorem \ref{randext} to convert the error guarantee in relative entropy to error guarantee in purified distance). This completes the proof.
\end{proof}

In the asymptotic and i.i.d. setting, we obtain the following corollary, using the technique of recycling the shared entanglement as elaborated in Theorem \ref{asympmeascomp}. This recovers the corresponding result in \cite{WildeHBH12}.
\begin{corollary}
Let $\ket{\Psi}_{RA_0C}$ be a quantum state and $\cN:\cL(\cH_{A_0}) \rightarrow \cL(\cH_{A}\otimes \cH_C)$ be a quantum measurement. Let $\cM: \cL(\cH_{A_0})\rightarrow \cL(\cH_A\otimes \cH_W)$ be a quantum measurement acting as $$\cM(\rho_{A_0}) = \sum_w \ketbra{w}_W\otimes M^{\dagger}_w\rho_{A_0}M_w$$ and $p(c | w)$ be a probability distribution for each $w$, such that $$\cN(\rho_{A_0}) = \sum_{c,w} p(c|w) \ketbra{c}_C\otimes M_w\rho_{A_0}M^{\dagger}_w.$$ Define the quantum state $$\Psi_{RBACW}\defeq \sum_{c,w} p(c|w) \ketbra{c}_C\otimes \ketbra{w}_W\otimes M^{\dagger}_w\Psi_{RBA_0}M_w.$$ For every $\eps \in (0,1/13)$ and integer $n \geq 1$, there exists a $(Q(n,\eps), S(n, \eps), O(\sqrt{n}), 13\eps)$- quantum measurement compression protocol without feedback for $\ket{\Psi}^{\otimes n}_{RA_0C}$ and $\cN^{\otimes n}$ such that 
$$\lim_{n\rightarrow \infty}\frac{1}{n} Q(n,\eps) = \condmutinf{R}{W}{B}, \lim_{n\rightarrow \infty}\frac{1}{n} S(n,\eps) = \mathrm{H}(W|BR) - \mathrm{H}(W|CBR).$$
\end{corollary}
\begin{proof}
The proof follows along the lines similar to Theorem \ref{asympmeascomp}, by applying the achievability proof given in Theorem \ref{theo:meascompfed} and using Fact \ref{dmaxequi}.
\end{proof}

\subsection*{Conclusion}

We have studied the problem of quantum measurement compression with quantum side information in the one-shot setting. Previously, this task had been studied only in the asymptotic and i.i.d. setting \cite{Winter04, WildeHBH12}. We have discussed the communication required to achieve such a task and the randomness cost of the protocol. These discussions are facilitated by a new formulation of the convex-split lemma which allows for a substantial reduction in the randomness cost. As a result, we obtain optimal rates of communication and randomness cost in the asymptotic and i.i.d. setting, obtaining a new optimal protocol for this task. 

Furthermore, we also obtain a new protocol for the important cryptographic primitive of strong randomness extraction in presence of quantum side information, using the convex-split lemma with limited randomness and obtain near optimal number of uniform bits (characterized by the conditional min-entropy \cite{Renner05}). An important question in this direction is if we can reduce the number of bits of initial randomness (the state of the art being much smaller, as obtained in \cite{DePVR12}). Through a composition of our protocol for randomness extraction and quantum measurement compression, we obtain a one-shot protocol for quantum measurement compression without feedback, which converges to the optimal rate in the asymptotic and i.i.d. setting \cite{WildeHBH12}.

An exciting problem is to use the one-shot quantum measurement compression results (with side information), for tasks such as one-shot purity distillation \cite{Devetak05,HorodeckiOSS03,HorodeckiOSSS05,DevetakKrovi07, DevetakHW08} or simulation of measurements on quantum states shared between Alice, Bob and Reference. Another important question is if it is possible to reduce the amount of shared entanglement in protocols that use convex-split and position based decoding techniques, potentially employing a fully quantum notion of pairwise independence or exploiting the ideas developed in \cite{ChaoRSV17}. Present formulation only applies in the classical-quantum setting with shared randomness, which does not cover all possible quantum information theoretic scenarios. Along similar lines, it is plausible that our techniques would lead to reduction in the amount of catalyst used in the works \cite{AnshuHJ17, BertaM17} for the randomness cost of resource destruction in resource theories. 

\subsection*{Acknowledgement} We thank Mark M. Wilde for introducing us to this problem and for many helpful comments. We also thank Eneet Kaur for helpful feedback on our manuscript. We thank Jaikumar Radhakrishnan for helpful discussions related to randomness extractors.
 
This work is supported by the Singapore Ministry of Education and the National Research Foundation, through the Tier 3 Grant ``Random numbers from quantum processes'' MOE2012-T3-1-009, NRF RF Award NRF-NRFF 2013-13 and VAJRA Grant, Department of Science and Technology, Government of India. Part of the work done when R. J. was visiting TIFR, Mumbai, India under the VAJRA Scheme of Government of India. Part of the work was done when A. A was visiting TIFR, Mumbai, India.

\bibliographystyle{ieeetr}
\bibliography{References}

\begin{thebibliography}{10}

\bibitem{Winter04}
A.~Winter, ````extrinsic'' and ``intrinsic'' data in quantum measurements:
  Asymptotic convex decomposition of positive operator valued measures,'' {\em
  Communications in Mathematical Physics}, vol.~244, no.~1, pp.~157--185, 2004.

\bibitem{MassarP00}
S.~Massar and S.~Popescu, ``Amount of information obtained by a quantum
  measurement,'' {\em Phys. Rev. A}, vol.~61, p.~062303, May 2000.

\bibitem{WinterM01}
A.~Winter and S.~Massar, ``Compression of quantum-measurement operations,''
  {\em Phys. Rev. A}, vol.~64, p.~012311, Jun 2001.

\bibitem{HsiehW15}
M.~H. Hsieh and S.~Watanabe, ``Source compression with a quantum helper,'' in
  {\em 2015 IEEE International Symposium on Information Theory (ISIT)},
  pp.~2762--2766, June 2015.

\bibitem{Devetak05}
I.~Devetak, ``Distillation of local purity from quantum states,'' {\em Phys.
  Rev. A}, vol.~71, p.~062303, Jun 2005.

\bibitem{HorodeckiOSS03}
M.~Horodecki, K.~Horodecki, P.~Horodecki, R.~Horodecki, J.~Oppenheim,
  A.~Sen(De), and U.~Sen, ``Local information as a resource in distributed
  quantum systems,'' {\em Phys. Rev. Lett.}, vol.~90, p.~100402, Mar 2003.

\bibitem{HorodeckiOSSS05}
M.~Horodecki, P.~Horodecki, R.~Horodecki, J.~Oppenheim, A.~Sen(De), U.~Sen, and
  B.~Synak-Radtke, ``Local versus nonlocal information in quantum-information
  theory: Formalism and phenomena,'' {\em Phys. Rev. A}, vol.~71, p.~062307,
  Jun 2005.

\bibitem{DevetakKrovi07}
H.~Krovi and I.~Devetak, ``Local purity distillation with bounded classical
  communication,'' {\em Phys. Rev. A}, vol.~76, p.~012321, Jul 2007.

\bibitem{DevetakHW08}
I.~Devetak, A.~W. Harrow, and A.~J. Winter, ``A resource framework for quantum
  shannon theory,'' {\em IEEE Transactions on Information Theory}, vol.~54,
  pp.~4587--4618, Oct 2008.

\bibitem{WildeHBH12}
M.~M. Wilde, P.~Hayden, F.~Buscemi, and M.-H. Hsieh, ``The
  information-theoretic costs of simulating quantum measurements,'' {\em
  Journal of Physics A: Mathematical and Theoretical}, vol.~45, no.~45,
  p.~453001, 2012.

\bibitem{Berta09}
M.~Berta, ``Single-shot quantum state merging,'' 2009.
\newblock Master's thesis, ETH Zurich.

\bibitem{Renner11}
M.~Berta, M.~Christandl, and R.~Renner, ``The {Q}uantum {R}everse {S}hannon
  {T}heorem based on one-shot information theory,'' {\em Commun. Math. Phys.},
  vol.~306, pp.~579--615, 2011.

\bibitem{horodecki07}
M.~Horodecki, J.~Oppenheim, and A.~Winter, ``Quantum state merging and negative
  information,'' {\em Communications in Mathematical Physics}, vol.~269,
  pp.~107--136, 2007.

\bibitem{Berta14}
M.~Berta, M.~Christandl, and D.~Touchette, ``Smooth entropy bounds on one-shot
  quantum state redistribution,'' {\em IEEE Transactions on Information
  Theory}, vol.~62, pp.~1425--1439, March 2016.

\bibitem{AnshuDJ14}
A.~Anshu, V.~K. Devabathini, and R.~Jain, ``Quantum communication using
  coherent rejection sampling,'' {\em Phys. Rev. Lett.}, vol.~119, p.~120506,
  Sep 2017.

\bibitem{AnshuJW17-1}
A.~Anshu, R.~Jain, and N.~A. Warsi, ``Smooth min-max relative entropy based
  bounds for one-shot classical and quantum state redistribution.''
  https://arxiv.org/abs/ arXiv:1702.02396, 2017.

\bibitem{Devatakyard}
I.~Devetak and J.~Yard, ``Exact cost of redistributing multipartite quantum
  states,'' {\em Phys. Rev. Lett.}, vol.~100, 2008.

\bibitem{YardD09}
J.~T. Yard and I.~Devetak, ``Optimal quantum source coding with quantum side
  information at the encoder and decoder.,'' {\em IEEE Transactions on
  Information Theory}, vol.~55, pp.~5339--5351, 2009.

\bibitem{RenesR12}
J.~M. Renes and R.~Renner, ``One-shot classical data compression with quantum
  side information and the distillation of common randomness or secret keys,''
  {\em IEEE Transactions on Information Theory}, vol.~58, pp.~1985--1991, March
  2012.

\bibitem{Renner05}
R.~Renner, ``Security of quantum key distribution.'' PhD Thesis, ETH Zurich,
  Diss. ETH No. 16242, arXiv:quant-ph/0512258, 2005.

\bibitem{KonigT08}
R.~T. Konig and B.~M. Terhal, ``The bounded-storage model in the presence of a
  quantum adversary,'' {\em IEEE Transactions on Information Theory}, vol.~54,
  pp.~749--762, Feb 2008.

\bibitem{Tashma09}
A.~Ta-Shma, ``Short seed extractors against quantum storage,'' in {\em
  Proceedings of the Forty-first Annual ACM Symposium on Theory of Computing},
  STOC '09, (New York, NY, USA), pp.~401--408, ACM, 2009.

\bibitem{TomRSS10}
M.~Tomamichel, R.~Renner, C.~Schaffner, and A.~Smith, ``Leftover hashing
  against quantum side information,'' in {\em 2010 IEEE International Symposium
  on Information Theory}, pp.~2703--2707, June 2010.

\bibitem{Hayashi11}
M.~Hayashi, ``Exponential decreasing rate of leaked information in universal
  random privacy amplification,'' {\em IEEE Transactions on Information
  Theory}, vol.~57, pp.~3989--4001, June 2011.

\bibitem{DePVR12}
A.~De, C.~Portmann, T.~Vidick, and R.~Renner, ``Trevisan's extractor in the
  presence of quantum side information,'' {\em SIAM Journal on Computing},
  vol.~41, no.~4, pp.~915--940, 2012.

\bibitem{Berta13}
M.~Berta, ``Quantum side information: Uncertainty relations, extractors,
  channel simulations.'' PhD Thesis, ETH Zurich, Diss. ETH No. 21180,
  arXiv:quant-ph/1310.4581, 2005.

\bibitem{BertaFW14}
M.~Berta, O.~Fawzi, and S.~Wehner, ``Quantum to classical randomness
  extractors,'' {\em IEEE Transactions on Information Theory}, vol.~60,
  pp.~1168--1192, Feb 2014.

\bibitem{BennettBCM95}
C.~H. Bennett, G.~Brassard, C.~Crepeau, and U.~M. Maurer, ``Generalized privacy
  amplification,'' {\em IEEE Transactions on Information Theory}, vol.~41,
  pp.~1915--1923, Nov 1995.

\bibitem{LeditzkyWD16}
F.~Leditzky, M.~M. Wilde, and N.~Datta, ``Strong converse theorems using
  r\'{e}nyi entropies,'' {\em Journal of Mathematical Physics}, vol.~57, no.~8,
  p.~082202, 2016.

\bibitem{AnshuJW17}
A.~Anshu, R.~Jain, and N.~A. Warsi, ``One shot entanglement assisted classical
  and quantum communication over noisy quantum channels: A hypothesis testing
  and convex split approach.'' https://arxiv.org/abs/1702.01940, 2017.

\bibitem{Wilde17a}
M.~M. Wilde, ``Position-based coding and convex splitting for private
  communication over quantum channels.'' https://arxiv.org/abs/1703.01733,
  2017.

\bibitem{AnshuJW17_cc}
A.~Anshu, R.~Jain, and N.~A. Warsi, ``A unified approach to source and message
  compression.'' https://arxiv.org/abs/1707.03619, 2017.

\bibitem{AhlswedeW02}
R.~Ahlswede and A.~Winter, ``Strong converse for identification via quantum
  channels,'' {\em IEEE Transactions on Information Theory}, vol.~48,
  pp.~569--579, Mar 2002.

\bibitem{Devetak05private}
I.~Devetak, ``The private classical capacity and quantum capacity of a quantum
  channel,'' {\em IEEE Transactions on Information Theory}, vol.~51,
  pp.~44--55, Jan 2005.

\bibitem{RadhaSW17}
J.~Radhakrishnan, P.~Sen, and N.~A. Warsi, ``One-shot private classical
  capacity of quantum wiretap channel: Based on one-shot quantum covering
  lemma.'' https://arxiv.org/abs/1703.01932, 2017.

\bibitem{Josza94}
R.~Jozsa, ``Fidelity for mixed quantum states,'' {\em Journal of Modern
  Optics}, vol.~41, no.~12, pp.~2315--2323, 1994.

\bibitem{uhlmann76}
A.~Uhlmann, ``The "transition probability" in the state space of a *-algebra,''
  {\em Rep. Math. Phys.}, vol.~9, pp.~273--279, 1976.

\bibitem{GilchristLN05}
A.~Gilchrist, N.~K. Langford, and M.~A. Nielsen, ``Distance measures to compare
  real and ideal quantum processes,'' {\em Phys. Rev. A}, vol.~71, p.~062310,
  Jun 2005.

\bibitem{Neumann32}
J.~V. Neumann, {\em Mathematische Grundlagen der Quantenmechanik}.
\newblock Berlin, Germany: Springer, 1932.

\bibitem{umegaki1954}
H.~Umegaki, ``Conditional expectation in an operator algebra, i,'' {\em Tohoku
  Math. J. (2)}, vol.~6, no.~2-3, pp.~177--181, 1954.

\bibitem{Datta09}
N.~Datta, ``Min- and max- relative entropies and a new entanglement monotone,''
  {\em IEEE Transactions on Information Theory}, vol.~55, pp.~2816--2826, 2009.

\bibitem{Jain:2009}
R.~Jain, J.~Radhakrishnan, and P.~Sen, ``A property of quantum relative entropy
  with an application to privacy in quantum communication,'' {\em J. ACM},
  vol.~56, pp.~33:1--33:32, Sept. 2009.

\bibitem{BuscemiD10}
F.~Buscemi and N.~Datta, ``The quantum capacity of channels with arbitrarily
  correlated noise,'' {\em IEEE Transactions on Information Theory}, vol.~56,
  pp.~1447--1460, 2010.

\bibitem{hayashinagaoka}
M.~Hayashi and H.~Nagaoka, ``A general formula for the classical capacity of a
  general quantum channel,'' in {\em Proceedings IEEE International Symposium
  on Information Theory,}, pp.~71--, 2002.

\bibitem{Tomamichel12}
M.~Tomamichel, ``A framework for non-asymptotic quantum information theory,''
  2012.
\newblock PhD Thesis, ETH Zurich.

\bibitem{FuchsG99}
C.~A. Fuchs and J.~van~de Graaf, ``Cryptographic distinguishability measures
  for quantum-mechanical states,'' {\em IEEE Transactions on Information
  Theory}, vol.~45, pp.~1216--1227, May 1999.

\bibitem{barnum96}
H.~Barnum, C.~M. Cave, C.~A. Fuch, R.~Jozsa, and B.~Schmacher, ``Noncommuting
  mixed states cannot be broadcast,'' {\em Phys. Rev. Lett.}, vol.~76, no.~15,
  pp.~2818--2821, 1996.

\bibitem{lindblad75}
G.~Lindblad, ``Completely positive maps and entropy inequalities,'' {\em
  Commun. Math. Phys.}, vol.~40, pp.~147--151, 1975.

\bibitem{Winter:1999}
A.~Winter, ``Coding theorem and strong converse for quantum channels.,'' {\em
  IEEE Transactions on Information Theory}, vol.~45, no.~7, pp.~2481--2485,
  1999.

\bibitem{Ogawa:2002}
T.~Ogawa and H.~Nagaoka, ``A new proof of the channel coding theorem via
  hypothesis testing in quantum information theory,'' in {\em Information
  Theory, 2002. Proceedings. 2002 IEEE International Symposium on}, pp.~73--,
  2002.

\bibitem{Jain:2003a}
R.~Jain, J.~Radhakrishnan, and P.~Sen, ``A lower bound for the bounded round
  quantum communication complexity of set disjointness,'' in {\em 44th Annual
  IEEE Symposium on Foundations of Computer Science, 2003. Proceedings.},
  pp.~220--229, Oct 2003.

\bibitem{CastelleHR78}
D.~Dacunha-Castelle, H.~Heyer, and B.~Roynette, ``Ecole d'{E}t{\'e} de
  {P}robabilit{\'e}s de {S}aint-{F}lour {VII},'' {\em Lecture Notes in
  Mathematics, Springer-Verlag}, vol.~678, 1978.

\bibitem{TomHay13}
M.~Tomamichel and M.~Hayashi, ``A hierarchy of information quantities for
  finite block length analysis of quantum tasks,'' {\em IEEE Transactions on
  Information Theory}, vol.~59, pp.~7693--7710, Nov 2013.

\bibitem{li2014}
K.~Li, ``Second-order asymptotics for quantum hypothesis testing,'' {\em Ann.
  Statist.}, vol.~42, pp.~171--189, 02 2014.

\bibitem{Lovettnotes}
S.~Lovett, ``Pairwise independent hash functions and applications,'' 2015.
\newblock http://cseweb.ucsd.edu/~slovett/teaching/SP15-CSE190/.

\bibitem{carlen}
E.~Carlen, ``Trace inequalities and quantum entropy: an introductory course.
  entropy and the quantum,'' {\em Contemp. Math.}, vol.~529, pp.~73--140, 2010.

\bibitem{Chebysev1852}
P.~Tchebychev, ``M\'{e}moire sur les nombres premiers,'' {\em Journal de
  math\'{e}matiques pures et appliqu\'{e}es}, vol.~1, pp.~366--390, 1852.

\bibitem{MillerShi16}
C.~A. Miller and Y.~Shi, ``Robust protocols for securely expanding randomness
  and distributing keys using untrusted quantum devices,'' {\em J. ACM},
  vol.~63, pp.~33:1--33:63, Oct. 2016.

\bibitem{RadhakrishnanT00}
J.~Radhakrishnan and A.~Ta-Shma, ``Bounds for dispersers, extractors, and
  depth-two superconcentrators,'' {\em SIAM Journal on Discrete Mathematics},
  vol.~13, no.~1, pp.~2--24, 2000.

\bibitem{ChaoRSV17}
R.~Chao, B.~W. Reichardt, C.~Sutherland, and T.~Vidick, ``Overlapping qubits.''
  https://arxiv.org/abs/1701.01062, 2017.

\bibitem{AnshuHJ17}
A.~Anshu, M.-H. Hsieh, and R.~Jain, ``Quantifying resource in catalytic
  resource theory.'' https://arxiv.org/abs/1708.00381, 2017.

\bibitem{BertaM17}
M.~Berta and C.~Majenz, ``Disentanglement cost of quantum states.''
  https://arxiv.org/abs/1708.00360, 2017.

\end{thebibliography}
\end{document}